\documentclass[lettersize,journal]{IEEEtran}
\usepackage{amsfonts}
\usepackage{times}
\usepackage{graphicx}
\usepackage{epstopdf}
\usepackage{latexsym}
\usepackage{dsfont}
\usepackage{amssymb}
\usepackage{amsmath}
\usepackage{cite}
\usepackage{verbatim}

\newcommand{\figref}[1]{{Fig.}~\ref{#1}}

\def\bb0{{\mathbb{0}}}

\def\bb{{\mathbf{b}}}

\def\bp{{\mathbf{p}}}

\def\b0{{\mathbf{0}}}

\def\bB{{\mathbf{B}}}
\def\bC{{\mathbf{C}}}
\def\bD{{\mathbf{D}}}

\def\bP{{\mathbf{P}}}

\def\bbC{{\mathbb{C}}}

\def\bbE{{\mathbb{E}}}

\def\sf0{{\mathsf{0}}}

\usepackage{graphicx}
\usepackage{epstopdf}
\usepackage{enumerate}
\usepackage{algorithm}
\usepackage{algorithm,algcompatible,amsmath}
\algnewcommand\INPUT{\item[\textbf{Input:}]}%
\algnewcommand\OUTPUT{\item[\textbf{Output:}]}%
\usepackage{amsmath}
\usepackage{mathrsfs}
\usepackage{float}
\usepackage{color}
\usepackage{makeidx}
\usepackage{bm}
\usepackage{cleveref}
\usepackage{url}
\usepackage{steinmetz}
\usepackage{varwidth}
\usepackage{soul}
\usepackage{bbm}
\usepackage{empheq}
\usepackage{mathtools}
\usepackage{cite}
\usepackage{balance}
\usepackage[scriptsize]{subfigure}

\def \rm {\mathrm}

\usepackage{amsfonts}
\usepackage{booktabs}
\usepackage{siunitx}

\usepackage{tikz}

\begin{document}
	
	\title{Vision Guided MIMO Radar Beamforming for Enhanced Vital Signs Detection in Crowds}
	
	\author{
		Shuaifeng~Jiang,
		~Ahmed~Alkhateeb,
            ~Daniel~W. Bliss,
            and~Yu~Rong
		\thanks{Y. R.  and D. W. B.  are with Center for Wireless Information Systems and Computational Architectures (WISCA) at the School of Electrical, Computer and Energy Engineering, Arizona State University, Tempe, AZ 85281 USA, e-mail: \{yrong5\}@asu.edu.}
		\thanks{S. J. and A. A. are with the Wireless Intelligence Lab at the School of Electrical, Computer and Energy Engineering, Arizona State University, e-mail: \{s.jiang, alkhateeb\}@asu.edu}
	}

	
	\maketitle
	
	\begin{abstract}
		Radar as a remote sensing technology has been used to analyze human activity for decades. Despite all the great features such as motion sensitivity, privacy preservation, penetrability, and more, radar has limited spatial degrees of freedom compared to optical sensors and thus makes it challenging to sense crowded environments without prior information. In this paper, we develop a novel dual-sensing system, in which a vision sensor is leveraged to guide digital beamforming in a multiple-input multiple-output (MIMO) radar. Also, we develop a calibration algorithm to align the two types of sensors and show that the calibrated dual system achieves about two centimeters precision in three-dimensional space within a field of view of $75^\circ$ by $65^\circ$ and for a range of two meters. Finally, we show that the proposed approach is capable of detecting the vital signs simultaneously for a group of closely spaced subjects, sitting and standing, in a cluttered environment, which highlights a promising direction for vital signs detection in realistic environments.
	\end{abstract}
	
	
	\section{Introduction}
	\IEEEPARstart{V}{ital} signs, such as heart and respiratory rates, provide critical insight into a person's well-being. Traditional methods of vital sign monitoring involve inconvenient uncomfortable wearable devices. To capture both heart and respiratory rates, two different sensors are required, e.g., a multi-led electrocardiography sensor which measures the heart’s electrical activity and a chest respiration belt which measures the air pressure variation according to the breathing of the subject. These standard monitoring wearables can typically measure a single object and do not apply to multi-subject scenarios. More convenient means of non-contact monitoring are then required to address these needs. Recent developments have successfully demonstrated radar technologies using single-device sensors for this purpose \cite{gu2016short}.

	There are two major competing remote sensing technologies for physiological parameter estimation: vision based approach and wireless based approach. A regular red, green and blue (RGB) camera can capture the color shift on the exposed facial skins under proper lighting conditions and the blood flow induced skin color tone change is used to extract blood pulse flow. The imaging based technology is often referred to as remote imaging plethysmography (RPPG) \cite{mcduff2015survey}. 
	The distinct drawback of using color cameras is the privacy issue. Thus, near-infrared (NIR) camera was investigated heavily based on the same RPPG principle for heart analysis because of the alleviated privacy concern and immunity of ambiance lighting change \cite{magdalena2018sparseppg}. However, the heartbeat pulse sensitivity at IR is significantly weaker compared to that of visible light (often green light) using RGB cameras.  
	More recently, state-of-the-art cameras based on time-of-flight (ToF) technology measure precise depth for respiratory monitoring via active IR projection to perceive depth \cite{kempfle2018respiration}. Using the depth measurements, the improved respiratory signal quality is observed from the depth pixel intensity change at area compared to the color pixel intensity change \cite{rongt2020respiration}.

	Radar is another promising technology for remote health monitoring. It enjoys all-weather features and raises no privacy issues compared to the vision counterpart. One unique advantage exists in radar is excellent material penetration capability, such as clothes and beddings, which is a desirable feature in daily health monitoring and patient care.
	Radar echos capture rich information of human body activity and inner organ movements. Physiological motion often shows repetitive yet time-varying patterns. The coherent transceiver in the radar system measures these minute motions based on the well-known micro-Doppler effect \cite{chen2006micro}.
Various types of radar systems were used to detect chest wall motion due to respiratory
and cardiac activities.
 Continuous-wave (CW) radars measure micromotion assuming no other sources of motion interference. When there are multiple sources of motion and/or multiple subjects, however, the performance of these narrow band systems degrades \cite{li2008random,rong2018harmonics}. Wideband radar systems, such as impulse ultra-wideband (UWB), frequency-modulated-CW (FMCW), and stepped-frequency-CW (SFCW), are able to distinguish targets at different ranges, but struggle to provide spatial positioning.  Utilizing an array of antennas provides increased degrees of freedom (DoF) compared to single-antenna systems \cite{bliss2003multiple}. This allows targets to be distinguished in both spatial angle and range. Advanced array processing techniques are available to boost radar sensing and imaging performance \cite{bliss2006mimo,rong2019radar,nosrati2019concurrent,feng2021multitarget}. Recently, the use of millimeter-wave (mmWave) and sub-terahertz (THz) frequency bands has gained popularity in remote vital signs monitoring \cite{ahmad2018vital,rong2021radar,rong2020cardiac}. With smaller wavelength, radar systems can be packed into a small form factor equipped with more antennas.

	\begin{figure}[t]
		\centering
		\includegraphics[width=1.\linewidth]{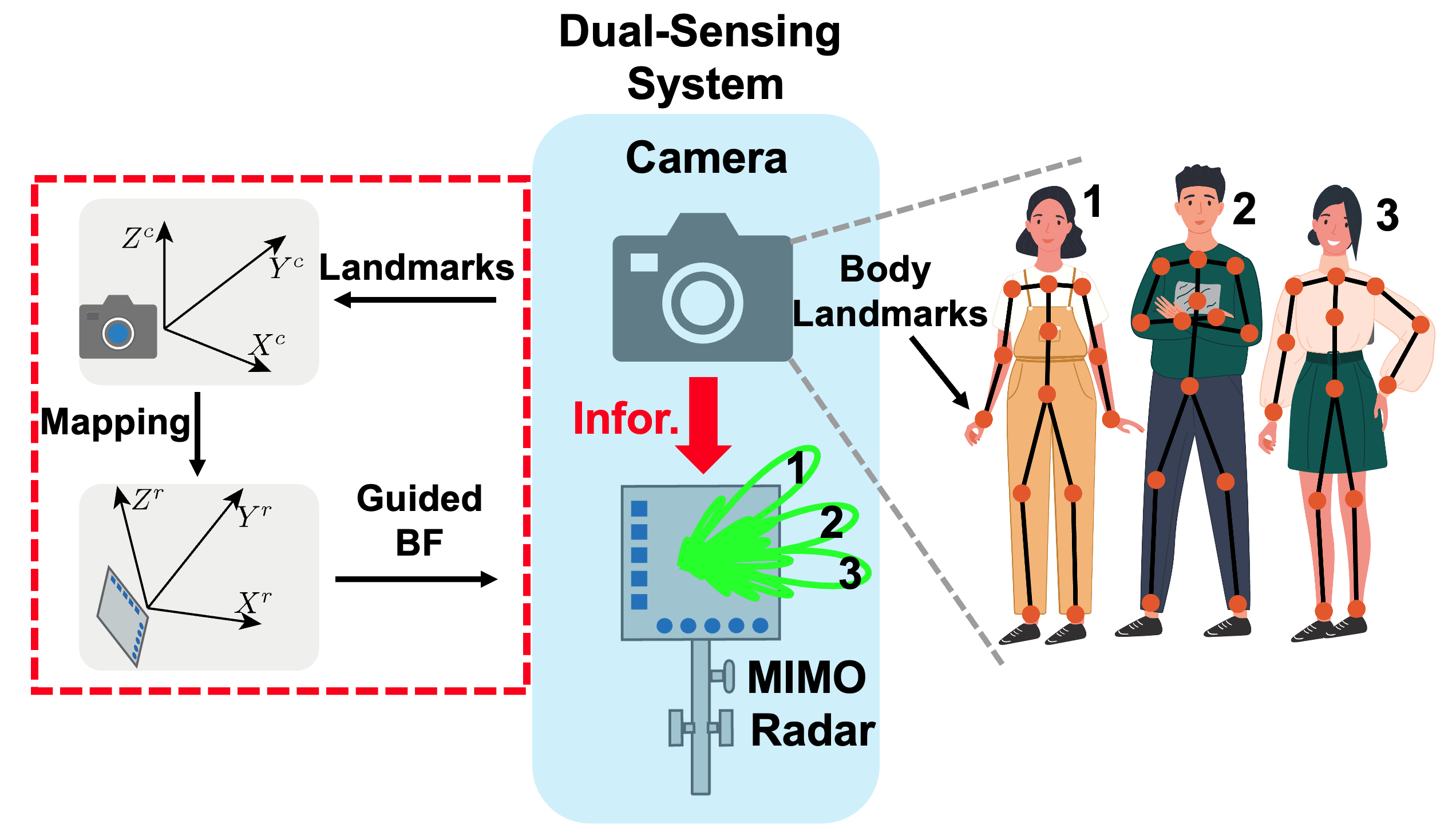}
		\caption{The graphical representation of the proposed concept - vision guided MIMO radar beamforming for VSD. The camera first estimates the 3-D coordinates of the torso landmarks of the subjects. These camera 3-D coordinates are then mapped to the radar coordinates to create high precision radar beamforming configuration.}
		\label{fig:Intro}
	\end{figure}
	
	\begin{table}[t]
		\caption{Notation Adopted in the Paper}
		\label{tbl:variable}
		\centering
		\begin{tabular}{|c|c|}
			\hline
			Variable & Description           \\ \hline
			$N$        & No. of Tx pulses in one SFCW frame   \\ 
			$f_{n}$   & Frequency of $n$-th pulse \\
			$f_{0}$  & Carrier frequency \\
			$\Delta f$ & Frequency step \\
			$T$  & Pulse duration \\
			$R_{0}$  & Target locations  \\
			$t, \tau$ & Slow-time, fast-time \\
			$d_{T}(t)$ & Target motion function of $t$ \\
			$\tau_{D}(t)$ & Motion induced time delay \\
			$c$ & Speed of light \\
			$s_{T}(\tau)$ & Tx signal\\
	        $S_{R}(t)$ & Baseband Rx signal at $R_{0}$  \\
			$g_{T}(x,y,z)$ & Tx aperture function \\
			$g_{R}(x,y,z)$ & Rx aperture function \\
			$x_{i,n_{i}}, y_{i,n_{i}}, z_{i,n_{i}}$  & Array element coordinate, where $i = T$ or $R$ \\
			$N_{i}$ & No. of Tx/Rx elements \\
			$\delta(.)$   & Dirac delta function \\
			$\theta,\phi$ & Azimuth and elevation angles \\
			$w_{n_{T},n_{R}}(\theta,\phi)$  & BF weight \\
			$k$ & Signal wavenumber \\
			$\gamma(x_{t},y_{t},x_{r},y_{r},k)$ & Scattering field \\
			$\Gamma(x_{t},y_{t},x_{r},y_{r},k)$ & Scattering field in spatial frequency domain \\
			$\emph{O}(x,y,z)$ & Imaging object function \\
			$k_{x_{t}},k_{y_{t}},k_{x_{r}},k_{y_{r}}$ & Fourier domain variables of $x_{t},y_{t},x_{r},y_{r}$ \\

			$\bD$, $\bC$ & Camera color image and point cloud\\
			$\bp^c=[x^c, y^c, z^c]^T$ & Camera coordinates of a point \\
			$\bp^r=[x^r, y^r, z^r]^T$ & Radar coordinates of a point \\
			$N_c$ & number of points in camera point cloud\\
			$\tilde{\bp}^c$, $\tilde{\bp}^r$ & coordinates of a point\\
			$\tilde{\bP}_c$, $\tilde{\bP}_r$ & coordinates of all
			measurements\\
			$B$, $\bB$ & Camera-radar mapping function/matrix \\
			\hline
		\end{tabular}
	\end{table}

	\subsection{Background}
	Sensor fusion concept for improved vital signs detection (VSD) was exploited in \cite{gu2013hybrid} by using a cellphone camera and CW radar to cancel random body motion. The cellphone camera was used to record large-scale body motions. The authors formulated a dual channel blind source separation problem. 
	
	Recently, a thermal camera and two independent CW radars (2.4 GHz) were used together in \cite{chian2022vital} for measuring vitals from multiple subjects. In their setup, the thermal camera only provided spatial identification and no depth information. Since the coordinates cannot be fused directly with the two non-coherent CW radar, lengthy signal processing was required to combine signals from the thermal camera and the two CW radars. The two independent radars were used only to expand radar coverage in azimuth and also created two independent channels with somewhat spatial diversity.

	A novel camera and radar fusion concept was presented in \cite{shokouhmand2022camera}. The system was constructed in a more natural way based on coordinate mapping. Their setup included a three-dimensional (3-D) depth camera to provide precise 3-D coordinates as a vision guidance and a mmWave multiple-input multiple-output (MIMO) to perform beamforming (BF) given the vision guidance. Critical information about vision coordinate and radar coordinate mapping performance was not given. But the subsequent sensing performance is directly related to the coordinate mapping accuracy.   More importantly, the full potential of vision guided radar for VSD has not been conveyed. Due to the limited array size, eight effective elements horizontally and two elements vertically, two human subjects with a minimum of 20 centimeters were demonstrated for VSD while 	prior works using similar mmWave radar without camera \cite{ahmad2018vital,xu2022simultaneous,xiong2021millimeter} reported success in separating two human subjects and estimating each subject's heart rate (HR) and respiratory rate (RR). 
	
	\subsection{Contributions}

	In this article, we are motivated to develop an accurate and automatic method to localize human subjects and thus estimate vital signs parameters from every individual at one time. We exploit the idea of vision-guided MIMO radar BF for VSD in a complex environment shown in Fig. \ref{fig:Intro}. 
	The dual-sensing system (DSS) consists of a 3-D depth camera and a low-power mmWave radar system with a multi-static MIMO array.  	20 transmit antennas (Tx) by 20 receive antennas (Rx) form total 400 virtual array elements.

	Human body landmarks is automatically recognized with the help of computer vision techniques. To map vision coordinate to radar coordinate, we develop a systematic approach. The setup involves a one-time calibration process by placing a corner reflector at different positions in the field of view (FoV) of DSS. The corner reflector locations are recorded by camera and radar. The 3-D coordinates of the corner reflector form the training data to solve an affine transformation with twelve DoFs that can describe the 3-D translation and rotation. After the calibration, the affine transformation coefficients are determined and used to construct the BF weights for radar.

	The main contributions of this paper can then be summarized as follows
	\begin{itemize}
		\item Develop a novel vision guided MIMO radar sensing system consisting of a MIMO imaging radar and a 3-D depth camera.
		\item Develop a systematic approach to align the radar and 3-D depth camera coordinates.
		\item Evaluate the localization performance of the DSS with corner reflector in 3-D.
		\item Develop a novel vision guided MIMO radar BF method for VSD and demonstrate with multiple subjects in the cluttered office environment.
	\end{itemize}
	
	\section{Motivation for Vision Guided Radar}
	\label{sec:motivation discussion}

\begin{figure}[t]
		\centering
		\resizebox{0.48\textwidth}{!}
		{\includegraphics{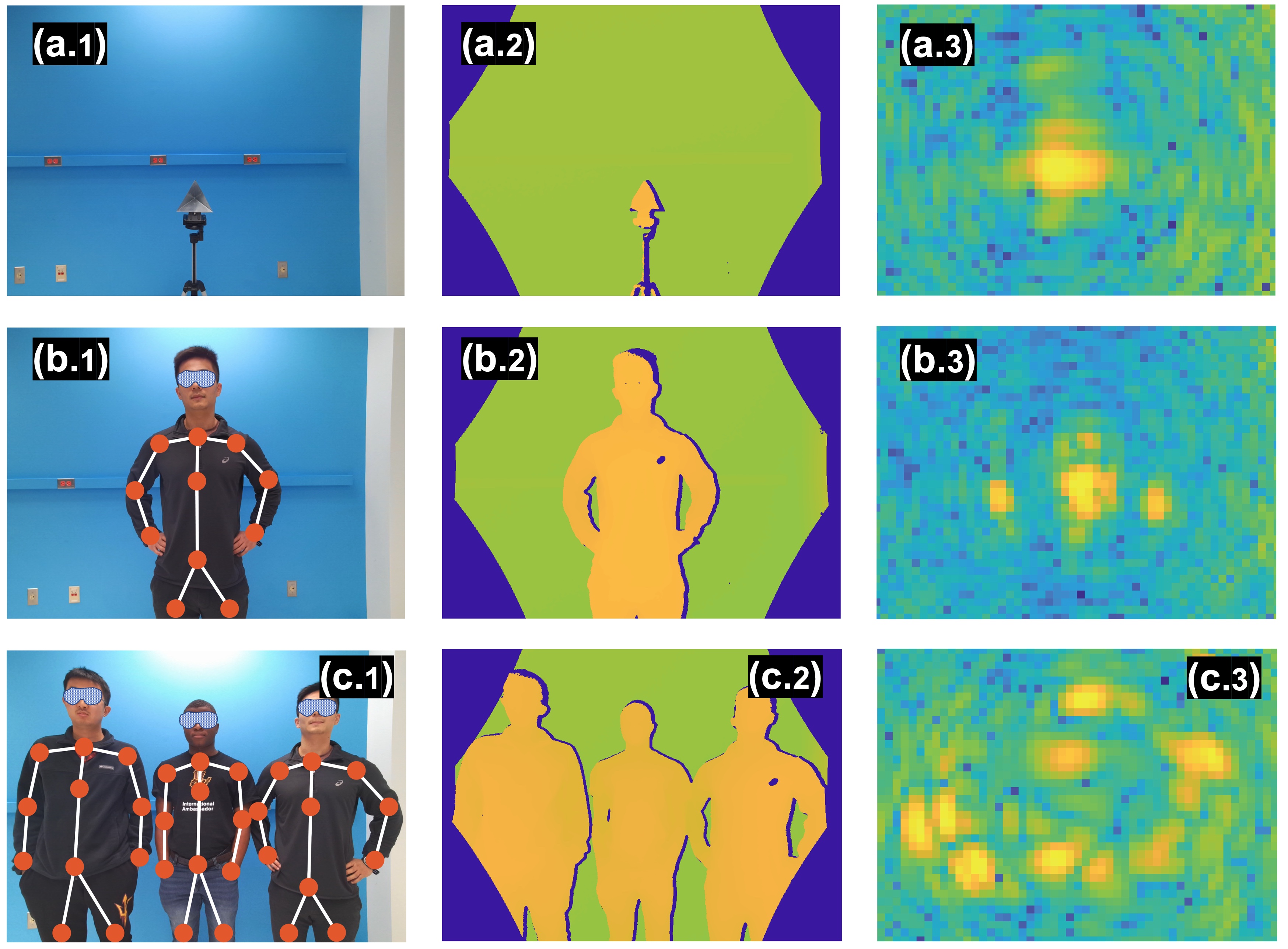}}
		\caption{Imaging performance comparison. Top row (a.1)-(a.3) displays RGB image, depth image (one slice of the 3-D depth clouds at the target distance), and radar image (one slice at the target distance); middle row displays the images from single subject with the human body landmarks highlighted in the RGB (b.1); bottom row displays the images from three human subjects located at the same imaging plane. \label{Plots::Motivation}}
        \end{figure}

\begin{figure}[t]
		\centering
		\resizebox{0.35\textwidth}{!}
		{\includegraphics{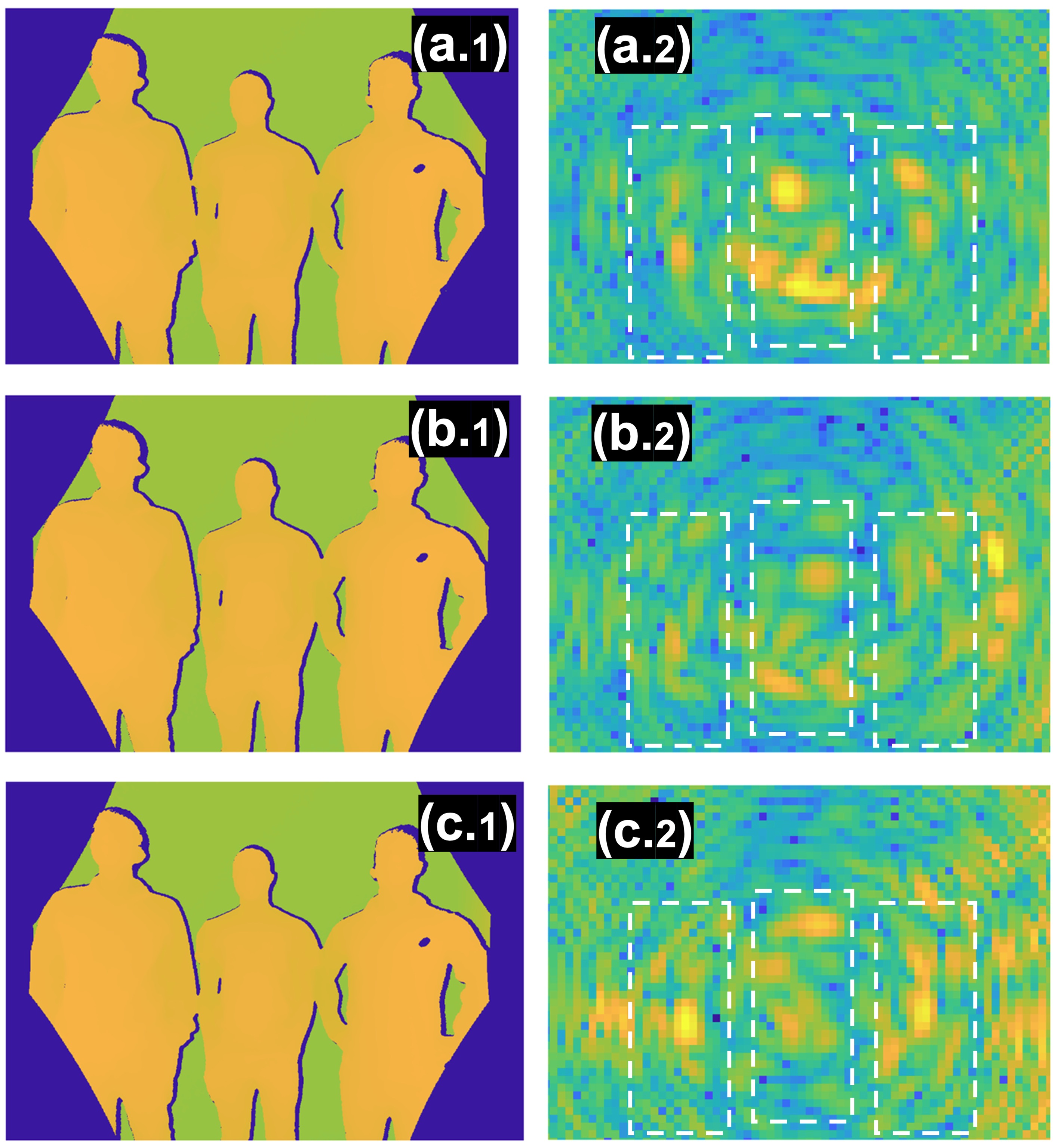}}
		\caption{Imaging stability in radar. Each row (a)-(c) represents one snapshot from a 5-second long dataset. Each column represents the depth imaging results (a.1)(b.1)(c.1) and the radar imaging results (a.2)(b.2)(c.2). Across the image frames (1-3), the depth imaging results are almost identical while the radar imaging results fluctuate quite a bit.\label{Plots::Stability}}
        \end{figure}

        Visual images, RGB and depth, from 3-D cameras provide far more visual contexts and textures compared to an advanced radar imaging system. Computer vision techniques and tools are already available and can be directly applied to visual images for human subject recognition \cite{turaga2008machine} and pose estimation \cite{sarafianos20163d}. These tasks are quite challenging to achieve with imaging radars. Here, and since the accuracy of the detection/recognition task is directly related to the imaging resolution, it is important to note that the imaging resolution is measured differently for cameras and radars.  
        For an optical device, the imaging resolution is given by the number of pixels. For a MIMO imaging radar, the imaging resolution is determined by the number of array elements and their locations. For example, in Fig. \ref{Plots::Motivation} and \ref{Plots::Stability}, the RGB image has 1920 pixels horizontally by 1080 vertically while the radar image is generated by a L-shape array with 20 physical elements horizontally by 20 vertically forming a 400-element virtual array as shown in Fig. \ref{Plots::MIMO Radar}(a)(c). Since the RGB and depth map are aligned, each pixel in the RGB image corresponds to a depth value in the depth image. Fig. \ref{Plots::Motivation} shows one snapshot RGB, depth and radar images from a corner reflector, single human subject and three human subject standing at the same imaging plane with respective to the sensor plane. Clearly, the visual images from the camera capture detailed structure of the imaging target for reflector and human subjects. Also, the object contour is revealed in the depth map. However, the radar images provide abstract information in the environment. Note that corner reflector is commonly used to calibrate electromagnetic wave devices by generating a larger radar cross-section. When imaging multiple subjects, the response pattern become complicated, which makes the automatically association of the radar return with each human subject almost impossible without a visual reference. 

        Imaging instability is another potential issue in radar imaging based sensing. This issue is demonstrated in Fig. \ref{Plots::Stability}. Three snapshots are randomly selected from a short 5-second recording. The depth images and the corresponding radar images are given in the left (a.1)-(c.1) and right columns (a.2)-(c.2). The imaging results from the radar fluctuates over time leading to ambiguity and unreliable localization. 
        Low emitting power, weaker reflection from human subjects compared to the clutterers, and sensitivity to involuntary body motion can explain this outcome.

	\section{MIMO Radar Signal Processing}
	\label{sec:signal model}
	In this section, we describe the adopted system and signal models for the MIMO radar.
	
	\subsection{SFCW Signal Model}
	
\begin{figure}
		\centerline{%
			\resizebox{0.5\textwidth}{!}{\includegraphics{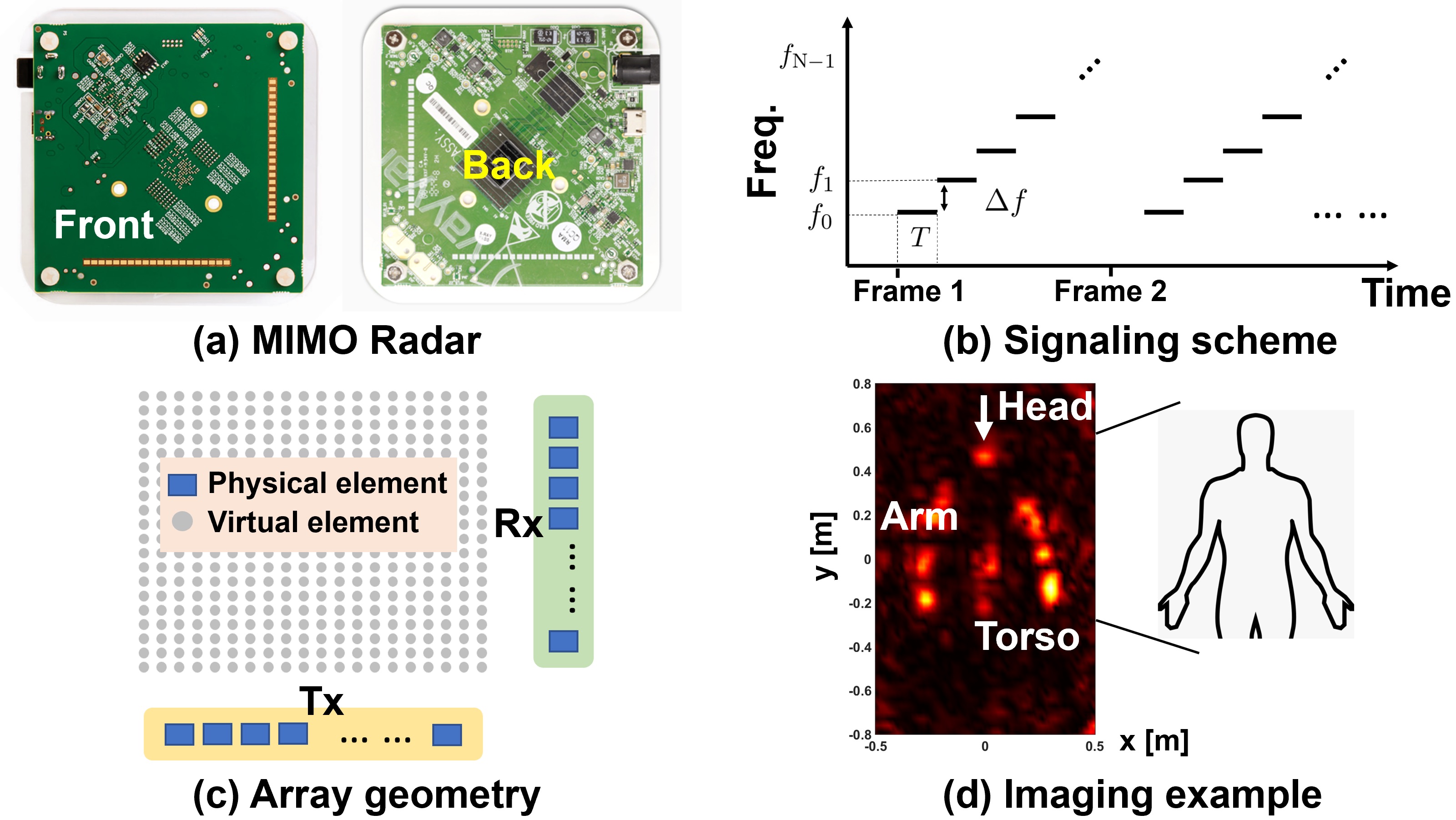}}%
		}
		\caption{MIMO imaging radar system. (a) Device figure, front and back; (b) SFCW signaling scheme; (c) 400-element virtual array; (d) radar imaging example of a human subject.}\label{Plots::MIMO Radar} 
	\end{figure}

	The 3-D radar imaging system deploys SFCW signaling and operates in the frequency domain rather than the time domain \cite{noon1996stepped}. The SFCW radar transmits a series of discrete narrow-band pulses in a stepwise manner to achieve a larger effective bandwidth. The modulated waveform consists of a group of $N$ coherent pulses with pulse duration $T$, and frequencies $f_{\rm n} = f_{0} + n \Delta f$. Assume that each SFCW waveform has $N$ pulses called one SFCW, frame and the center frequency of the first pulse is $f_{0}$, as illustrated in Fig. \ref{Plots::MIMO Radar}(b).
	
	One transmitted SFCW frame is represented as a sum of $N$ windowed narrow band signals, 
	\begin{equation}
	s_{T}(\tau) = \frac{1}{\sqrt{T}} \, \sum_{n} \, {\mathrm{rect}} \, \big ( \frac{\tau - n T}{T} \big ) \,\, {\rm e}^{{j}2\pi(f_{0} + n \Delta f)\tau} .
	\end{equation}
	
	The backscattered SFCW frame in the baseband is modeled by concatenating the down-converted received pulses. The received pulse is an attenuated and delayed version of the transmitted pulse at some distance. The range profile is obtained by performing inverse Fourier transform of the fast frequency samples. The slow-time signal at the target range is approximated as by ignoring other constant terms \cite{rong2022vital}, 
	\begin{align}
	S_{R}(t) \approx e^{{j}2 \pi f_{0} \tau_{\rm D}(t)},   \label{Eqn::sig}
	\end{align}  
	where $\tau_{\rm D}(t) = 2 R_{T}(t)/c$ denotes a slowly time-varying delay due to target motion $R_{T}(t)$. $c$ denotes the speed of light.

	\subsection{Time Division based MIMO Configuration}
	
	The SFCW signal frames were transmitted in a time division multiplexing (TDM) MIMO configuration by transmitting sequentially through the Tx antennas while the Rx antennas are acquiring the data all the time. 
	
	The MIMO virtual array is obtained by convolving the Tx and the Rx aperture functions $g_{T}(x,y,z)$, $g_{R}(x,y,z)$. For convenience, the last dependent variable $z$ is dropped since the Tx and the Rx arrays lie on the same $z=z_{a}$ plane. Thus, for the transmit array, the aperture illumination function is defined as
	\begin{equation}
	g_{T}(x,y) = \sum_{n_{T}=1}^{N_{T}} \delta(x-x_{T,n_{T}}, y-y_{T,n_{T}}).
	\end{equation}
	For the receiver array, the aperture illumination function is defined as 
	\begin{equation}
	g_{R}(x,y) = \sum_{n_{R}=1}^{N_{R}} \delta(x-x_{R,n_{R}}, y-y_{R,n_{R}}).
	\end{equation}
	The virtual array is given by 
	\begin{equation}
	\begin{split}
	g(x,y) =  & \big  ( g_{T} \ast g_{R} \big)(x,y) = \sum_{n_{T}=1}^{N_{T}} \sum_{n_{R}=1}^{N_{R}} \\ &\delta \big(x-(x_{T,n_{T}} + x_{R,n_{R}}),\,\, y-(y_{T,n_{T}} + y_{R,n_{R}}) \big )
	\end{split}
	\end{equation}
	where $x$ and $y$ together form coordinates within a 2-D plane, and $\delta(\cdot)$ denotes the Dirac delta function. Two linearly separated Tx antennas with spacing $\lambda$ and receive (Rx) array with spacing $\lambda$ form an $N_{T} \times N_{R}$ (400-element) virtual array as shown in Fig. \ref{Plots::MIMO Radar}(c).
	
	\subsection{Spatial Beamforming}
	
	BF is applied to isolate multiple individuals and increases the received signal-to-noise ratio (SNR). For a multiple Rx antenna system, the signal model in Eqn. (\ref{Eqn::sig}) will accordingly have an additional phase shift representing the phase progression across the array. For a target at an azimuth angle $\theta$ and an elevation angle $\phi$, the signal received from the $m$-th frame at the virtual element corresponding to the $n_{T}$-th Tx antenna and the $n_{R}$-th Rx antenna,
	\begin{align}
	s_{R}(t| n_{T},n_{R}) &= e^{{\rm j} 2 \pi f_{0} \tau_{\rm D}(t)}  e^{j2\pi \frac{  {\sin}(\theta) D_{x} + {\cos}(\phi) D_{y} }{\lambda}} \\
        \begin{split}
       &= e^{{\rm j} 2 \pi f_{0} \tau_{\rm D}(t)} \\ 
       &\times e^{j2\pi  \big( {\sin}(\theta)(n_{T}-1) + {\cos}(\phi)(n_{R}-1) \big)},\label{Eqn::+Angle}
       \end{split}
	\end{align}
    where $D_{x} = x_{T,n_{T}} + x_{R,n_{R}}$ and $D_{y} = y_{T,n_{T}} + y_{R,n_{R}}$. $\lambda$ is the wavelength. From Eqn. (6) to (7), the assumption of half-wavelength element spacing is invoked. The output of the BF is, 
	\begin{equation}
	\begin{split}
	s_{R}(t) &= \sum_{n_{T}} \sum_{n_{R}} s_{R}(t | n_{T},n_{R})  \,\, w^{\ast}_{n_{T},n_{R}}(\theta,\phi),  \\
	&= N_{T} \,\, N_{R} \,\, e^{{\rm j} 2 \pi f_{0} \tau_{\rm D}(t)} 
	\end{split}
	\end{equation}
	where $w_{n_{T},n_{R}}(\theta,\phi) = e^{j2\pi \big( (n_{T}-1){\sin}\theta + (n_{R}-1){\cos}\phi \big)}$.
	
	
	\subsection{3-D Radar Imaging}
	\label{sec:imaging}
	
	Imaging an object in a reflection arrangement is understood as the process of correlating the information of the reflected scattered field out of the object from multiple angles of view in order to find a description of the object function. In the imaging arrangement, Tx array and Rx array are placed at different positions within the $z=z_{z}$ plane. The collected data include all possible combinations of Tx and Rx at all frequencies of operation.
	
	The scattered field $\gamma(x_{t},y_{t},x_{r},y_{r},k)$ can be written as a volume integration as, 
	\begin{equation}
	\begin{split}
	\gamma(x_{t},y_{t},x_{r},y_{r},k) &= \int_{x}\int_{y}\int_{z} dx \, dy \, dz \,\, \emph{O}(x,y,z) \\
	&\times \,\, e^{-j k \sqrt{  (x_{t} - x)^{2} + (y_{t} - y)^{2} + (z_{a} - z)^{2}  }  } \\
	&\times \,\, e^{-j k \sqrt{(x_{t} - x)^{2} + (y_{t} - y)^{2} + (z_{a} - z)^{2}}}
	\end{split} \label{Eqn::Image1}
	\end{equation}
	
	The imaged object $\emph{O}(x,y,z)$ is reconstructed by solving the above inverse problem.
	The imaging reconstruction problem is solved based on spatial frequency domain representation \cite{nikolova2017introduction}. After taking 4-D Fourier transform on both sides of Eqn. \eqref{Eqn::Image1} with respect to $x_{t},y_{t},x_{r},y_{r}$, a more convenient representation, which matches the MIMO SFCW radar data recording in the frequency domain, is given as,
	\begin{equation}
	\gamma(x_{t},y_{t},x_{r},y_{r},k) = \emph{F}^{-1} \big \{ \Gamma(k_{x_{t}},k_{y_{t}},k_{x_{r}},k_{y_{r}},k) \big \}. \label{Eqn::Image2}
	\end{equation}
	$k_{x_{t}},k_{y_{t}},k_{x_{r}},k_{y_{r}}$ are the Fourier domain variables corresponding to $x_{t},y_{t},x_{r},y_{r}$.
	Plugging Eqn. \eqref{Eqn::Image1} into Eqn. \eqref{Eqn::Image2} and through some mathematical manipulations, the reconstructed object function using spatial frequency domain representation is hence expressed as,
	\begin{equation}
	\hat{\emph{O}}(x,y,z) = \emph{F}^{-1} \big \{  \Gamma(k_{x_{t}},k_{y_{t}},k_{x_{r}},k_{y_{r}},k) \times e^{-j k_{z} z_{a} }   \big \},
	\end{equation}
	where 
	\begin{equation}
	k_{z} = \sqrt{k^{2} - k^{2}_{x_{t}} - k^{2}_{y_{t}}} + \sqrt{k^{2} - k^{2}_{x_{r}} - q^{2}_{y_{r}}}.
	\end{equation}

	\section{Camera-Radar Dual Sensing System}
	Vision is expected to be more accurate than radar and the latter has some fundamental issues to localize multiple closely spaced human targets as discussed in the previous sections. This motivates us to develop a new technique based on vision-guided radar beamforming for sensing purposes. In this section, we describe the proposed camera-radar dual sensing system system.
	\subsection{Camera-Radar Dual Sensing System (DSS): The Testbed}
\begin{figure}[t]
		\centering
		\includegraphics[width=0.8\linewidth]{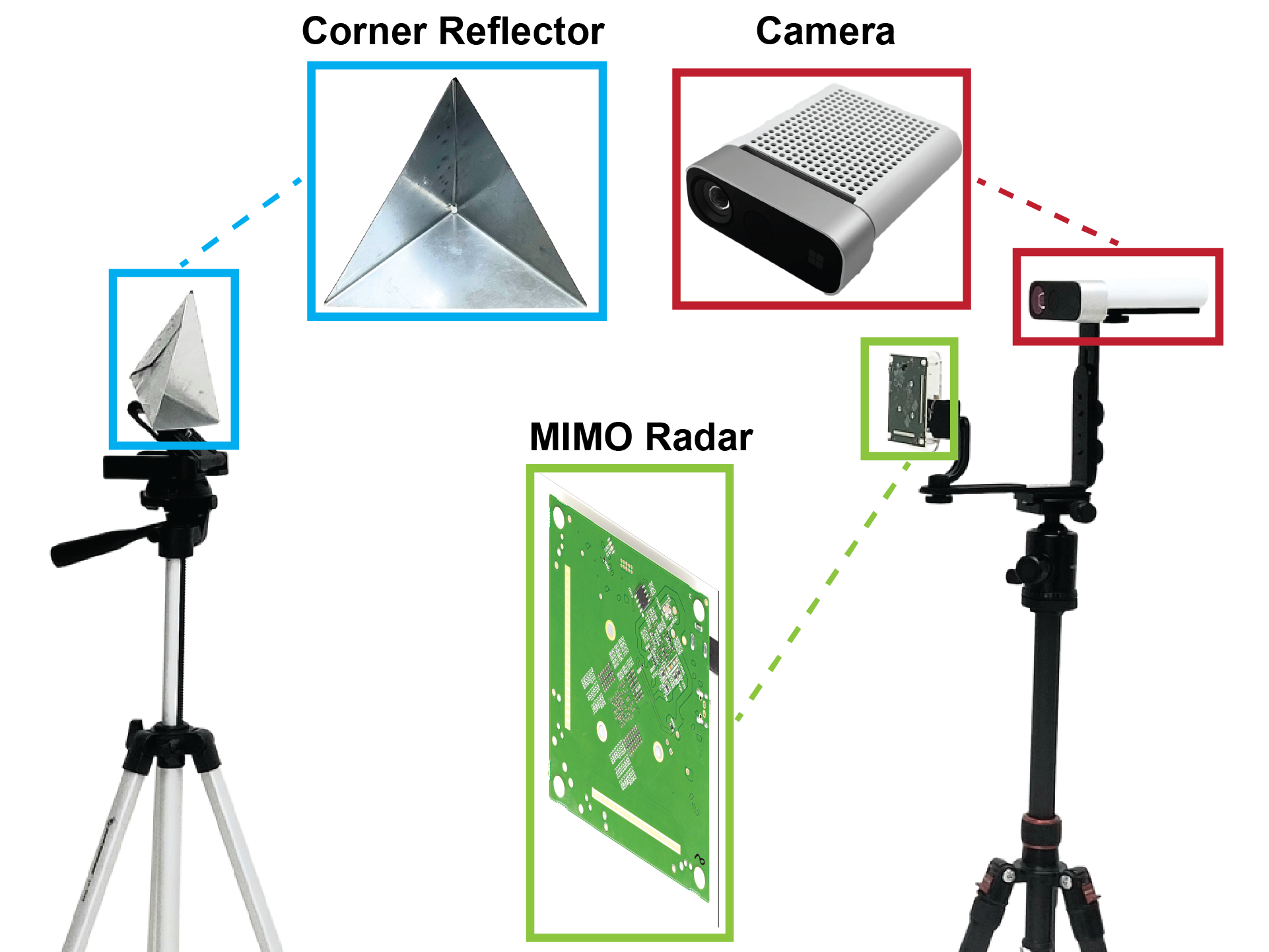}
		\caption{Snapshot of calibration setup to align camera and radar by varying a corner reflector in the FoV of the DSS.}
		\label{fig:testbed}
	\end{figure}

	The DSS is shown in \figref{fig:testbed}.
	The 3-D depth camera, Azure Kinec DK \cite{bamji2018impixel, azureDK_spec}, features a color camera and an advanced depth sensor using ToF technology.
	The color camera is configured to $1280$ by $720$ resolution with a FoV of $90^{\circ}$ by $59^{\circ}$. The depth camera is configured to the ``Narrow FOV unbinned" mode with a FoV of $75^{\circ}$ by $65^{\circ}$. The operating range of the depth camera is 0.3 to 3.86 meters. The FoV of the system is defined as the intersection of the FoVs of its color camera and depth sensor. The operations of the color camera and the depth sensor are synchronized in time with a frame rate of 30 frames/second (FPS). The key parameters of this camera system are summarized in Table \ref{tb:camera_params}.

	The mmWave imaging radar employs $N_T=20$ transmit antennas and $N_R =20$ receive antennas and operates in the frequency domain with the SFCW signaling. The frequency sweep contains 64 samples and covers 62 to 66 GHz with a 4 GHz bandwidth. The maximum detecting range is 2.34 meters. The IF resolution bandwidth of the radar is set to 20 kHz (related to integration time). These radar parameters deliver an imaging frame rate of about 18 FPS, which is sufficient for processing human vital signs. The maximum emitting power of the radar is about -10 dBm. The key parameters of the radar are summarized in Table \ref{tb:radar_params}. 
	
	\begin{table}[!t]
		\caption{3-D Camera System Parameters}
		\label{tb:camera_params}
		\centering
		\begin{tabular}{ll}
			\toprule
			- Camera System Configuration - & \\
			Color camera resolution &  $1280 \times 720$\\
			Color camera FoV & $90^{\circ} \times 59^{\circ}$\\
			Depth sensor mode & NFOV unbinned\\
			Depth sensor FoV & $75^{\circ} \times 65^{\circ}$ \\
			Frame rate & 30 FPS \\
			Operating range & 0.5-3.86 m\\
			\toprule
			- Point Cloud Accuracy - & \\
			$x$-axis & 2.75 cm\\
			$z$-axis & 2.68 cm\\
			$y$-axis & $< 1.5$ cm\\
			\bottomrule
		\end{tabular}
	\end{table}
	
	\begin{table}[!t]
		\caption{3-D mmWave imaging Radar Parameters}
		\label{tb:radar_params}
		\centering
		\begin{tabular}{ll}
			\toprule
			- Radar Configuration - & \\
			Start frequency & 62 GHz\\ 
			End frequency & 66 GHz\\ 
			Effective bandwidth & 4 GHz\\
			IF resolution bandwidth & 20 kHz\\
			Frame rate & 18 FPS\\
			Detecting range & 2.34 meters \\
			Max emitting power & -10 dBm \\
			Number of transmit antenna ($N_T$) & $20$\\
			Number of receive antenna ($N_R$) & $20$\\
			Number of samples per frequency sweep & 64\\
			\toprule
			- Radar Resolution at 0.5 meter boresight - & \\
			$x$-axis & 5 cm\\
			$z$-axis & 5 cm\\
			$y$-axis & 3.75 cm\\
			\bottomrule
		\end{tabular}
	\end{table}

	\subsection{Camera 3-D Point Cloud}
	The DK provides a 3-D point cloud $\bC$ for all the pixels in the color image that lie in the common FoV of the color camera and the depth sensor. Let $\bD$ and $\bC$ denote the RGB image and the 3-D point cloud captured by the camera system in one frame. With $N_c$ denoting the number of pixels, the 3-D point cloud can be written as $\bC=\{\bp^c_1, \hdots, \bp^c_{N_c}\}$, where $\bp^c=[x^c, y^c, z^c]^T$ are the horizontal, vertical, and depth coordinates of each pixel. 
	
\begin{figure}[t]
		\centering
		\includegraphics[width=0.6\linewidth]{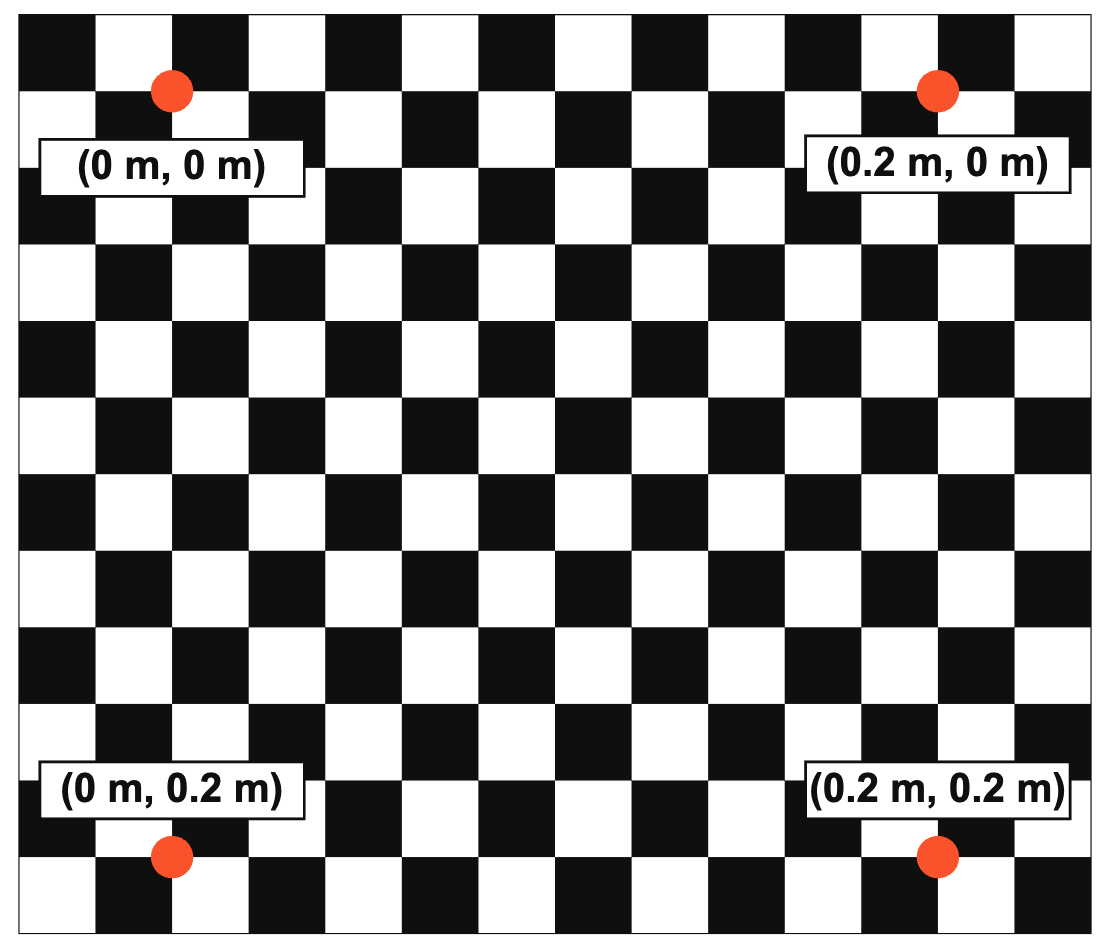}
		\caption{Use of checkboard with four vertices labeled to evaluate the depth point cloud accuracy in x/y-axis. \label{fig:checkboard}}
	\end{figure}

	According to the hardware specifications \cite{azureDK_spec}, the depth error of the Azure Kinect DK camera is less than 11 mm plus $0.1\%$ of distance without multi-path interference. This indicates that the mean absolute error (MAE) along the z-axis (depth) of $\bC$ is smaller than $1.5$ centimeter within $3.86$ meters.
	A checkboard is used to evaluate the point cloud accuracy in $\bC$ along the $x$ and $y$ axis. The checkboard is placed orthogonal to the camera z-axis and its checkboard grids are parallel to $x$ and $y$ axes. 
    Four vertices on the checkboard are selected as anchor points for evaluation. The anchors points are labeled as orange markers in Fig. \ref{fig:checkboard}. The distance among the four anchors are 20 centimeters vertically and horizontally. 
    For accuracy evaluation, the relative distance among anchors are estimated from $\bC$ as $(|x^c_2 - x^c_1| + |x^c_4 - x^c_3|) / 2$ and $(|y^c_2 - y^c_1| + |y^c_4 - y^c_3|) / 2$, where $x^c_i$ and $y^c_i$ are the anchor coordinates from $\bC$. 
    The MAE in $x$-axis and $y$-axis are $2.75$ cm and $2.68$ cm. The results are obtained by placing the checkboard at eight different positions.

	\subsection{MIMO Radar 3-D Spatial Resolution}
	The range resolution of the radar is 3.75 centimeters at the boresight given the 4 GHz signaling bandwidth. The azimuth and elevation angular resolution are determined by the antenna array layout and aperture size. Given the L-shape array in Fig. \ref{Plots::MIMO Radar}(c), the angular resolution in x-axis and y-axis are the same. The angular resolution $\Delta \theta$ is described by
	\begin{align}
	\Delta \theta &= \frac{\lambda}{N_T d_{x} \rm{cos}(\theta)}
	\end{align}
	where $d_{x}$ denotes the array element spacing in azimuth. Given $N_T = 20$ and $d_{x} = \lambda /2$, the best azimuth angular resolution occurs at the boresight, which is about $5.7^\circ$. The resulting x/y-axis spatial resolution in the Cartesian coordinate system at 0.5 meters is $5$ cm.

	\subsection{Camera-Radar Alignment}
	The goal of the camera-radar alignment is to estimate a transformation function $B: \bp^c \rightarrow \bp^r$ that can map a target's camera coordinate to its radar coordinate, in which $\bp^c = (x^c, y^c, z^c)$ and $\bp^r = (x^r, y^r, z^r)$ denote the Cartesian coordinates of an ideal point target in the camera and radar coordinate systems. Mathematically, the optimal transformation function $B^\star$ satisfies
	\begin{equation}\label{eq:trans}
	B^\star = \underset{B}{\arg\max}\, \bbE_{p}\left\{ \left\| \bp^r - B(\bp^c) \right \|_2^2  \right\},
	\end{equation}
	where $p$ is a random point in the camera-radar FoV.

	The transformation between the two coordinate systems can be modeled by an affine transformation, which can be used to describe the translation, reflection, rotation, and scaling transformations. Let $\tilde{\bp}^c = [x^c, y^c, z^c, 1]^T$ and $\tilde{\bp}^r = [\hat{x}^r, y^r, z^r, 1]^T$ denote the coordinates in the DSS. The mapping $\tilde{\bp}^r$ is given as
	\begin{equation}\label{eq:affine}
	\tilde{\bp}^r  = \bB \tilde{\bp}^c,
	\end{equation}
	where $\bB\in\bbC^{4 \times 4}$ is the affine transformation matrix. $\bB$ can be written as
	\begin{equation}
	\bB=
	\begin{bmatrix}
	t_{1,1} & t_{1,2} & t_{1,3}  & t_{1,4}\\
	t_{2,1} & t_{2,2} & t_{2,3}  & t_{2,4}\\
	t_{3,1} & t_{3,2} & t_{3,3}  & t_{3,4}\\
	0       &       0 & 0        & 1      \\
	\end{bmatrix}.
	\end{equation}

	To estimate $\bB$, the DSS needs $K_{\mathrm{tr}}$ different measurements to collect the training data at $K_{\mathrm{tr}}$ different positions. 
     Let $\tilde{\bp}^c_p$ and $\tilde{\bp}^c_p$ denote the coordinates of the $p$-th ($p \in \{1,\hdots, K_{\mathrm{tr}}\}$) position. Combining $K_{\mathrm{tr}}$ measurements Eqn. \eqref{eq:affine} can be re-written as
	\begin{equation}\label{eq:affine_all}
	\tilde{\bP}_r  = \bB \tilde{\bP}_c,
	\end{equation}
	where $\tilde{\bP}_r = [\tilde{\bp}^r_1, \hdots, \tilde{\bp}^r_{K_{\mathrm{tr}}}]$ and $\tilde{\bP}_c = [\tilde{\bp}^c_1, \hdots, \tilde{\bp}^c_{K_{\mathrm{tr}}}]$ concatenate the coordinates of all the $K_{\mathrm{tr}}$ targets. Substituting the $\bB$ into Eqn. \eqref{eq:trans}, the optimal $\bB$ can be estimated by solving the following optimization problem:
	\begin{equation}\label{eq:opt_aln_f}
	\underset{\bB}{\arg\min}\,  \left\| \tilde{\bP}_r -  \bB \tilde{\bP}_c \right\|_F^2  ,
	\end{equation}
	where the $\|\cdot\|_F$ denotes the Frobenius norm of a matrix. The least square estimated $\bB$ can be obtained by
	\begin{equation}\label{eq:affine_sol}
	\hat{\bB} = \tilde{\bP}_r \tilde{\bP}_c^\dagger, 
	\end{equation}
	where $\tilde{\bP}_c^\dagger = \tilde{\bP}_c (\tilde{\bP}_c \tilde{\bP}_c^T)^{-1}$ denote the pseudo inverse of $\tilde{\bP}_c$. $K_{\mathrm{tr}} \geq 4$ is required for estimating $\bB$ to ensure that $\tilde{\bP}_c$ is full row rank. The estimated radar coordinates $\hat{\bp}^r$ is obtained,
	\begin{equation}\label{eq:mapping}
	\hat{\bp}^r  = \hat{\bB} \tilde{\bp}^c.
	\end{equation}

    \subsection{Alignment Performance Evaluation}
    A highly reflective object is used in the calibration process. 
    Each measurement position contains the RGB image, depth image, and radar image as shown in Fig. \ref{Plots::Motivation}(a.1)-(a.3). The RGB pixel capturing the center of the corner reflector is manually selected. The corresponding pixel in the depth point cloud $\bC$ givens the camera coordinates $\bp^c$. 
    The peak location of the energy cluster in the radar image in Fig. \ref{Plots::Motivation}(a.3) is selected as the radar coordinate of the corner reflector center, $\bp^r$.
    
    Totally, 33 pairs of camera and radar coordinates are collected by placing the corner reflector at various locations in the FoV of the DSS.
    Among them $K_{\rm{tr}}=19$ measurements are used as the training data to estimate $\hat{\bB}$ in Eqn. \eqref{eq:affine_sol}. For evaluation of the calibrated DSS, the MAE along the $x$, $y$, and $z$ axis is calculated using $K_{\rm{test}}=14$ testing measurements. 
	The MAE on the training and testing is summarized in Table \ref{tb:camera_radar_mae}. The DSS achieves centimeter-level MAE performance on testing results. Our method provides sufficient camera-radar alignment accuracy for the DSS, and thus can be utilized to realize the vision guided radar sensing task.
	
	\begin{table}[]\caption{The camera-radar alignment mean absolute error obtained on the training and test datasets}\label{tb:camera_radar_mae}
		\centering
		\begin{tabular}{|l|l|l|l|l|}
			\hline
			MAE (centimeter)       & $x$   & $y$  & $z$    &Average\\ \hline
			Affine (Train)         & 1.07  & 2.20  & 0.80  &1.36\\ \hline
			Affine (Test)          & 1.82  & 3.85  & 0.49  &2.05\\ \hline
		\end{tabular}
	\end{table}

	\section{Vital Signs Extraction}
	
	A pre-trained pose estimation model \cite{Pose} is deployed to automatically estimate torso landmarks from single or multiple subjects. The target area for VSD is chosen as the center point of chest, which is interpolated based on the torso landmarks, as displayed in Fig. \ref{fig:Intro}. The mapping function transforms the target position $x^c$, $y^c$, and $z^c$ in the camera coordinate to the radar coordinate $\hat{x}^r$, $\hat{y}^r$, and $\hat{z}^r$ in Eqn. \eqref{eq:mapping}.
	The perspective angles $\theta_{0}$ and $\phi_{0}$ for VSD with respect to the boresight of the DSS are then calculated, 
	\begin{align}
	\theta_{0} &= {\mathrm{arctan}} \, \Big [ \frac{\hat{x}^r}{R_{0}} \Big ] \\
	\phi_{0} &= {\mathrm{arctan}} \, \Big [  \frac{\hat{y}^r}{R_{0}} \Big ],
	\end{align}
	where $R_{0}$ denotes the radial distance from the sensor to the human subject, given by $R_{0} = \sqrt{(\hat{x}^r)^2 + (\hat{y}^r)^2 + (\hat{z}^r)^2}$.
	The spatial channels are constructively combined to enhance SNR given a weak emitting power at maximum -10 dBm. The radial distance $R_{0}$ determines the range bin for VSD. 
	
	The phase component at $R_{0}$ is extracted by taking the unwrapping function and angle function on Eqn. (\ref{Eqn::sig}),
	\begin{equation}
	\Psi(t) = {\mathrm{unwarp}} \, \Big [ S_{R}(t) \Big ].
	\end{equation}
	The RR and HR are estimated by locating the spectral peaks in the relevant frequency regions. Prior to taking FFT on the phase signal, a high pass filter is applied to remove the DC component and the phase differential method \cite{ahmad2018vital} is also applied to emphasize weak heartbeat signal,
	\begin{equation}
	\psi(t) = \frac{\delta \Psi(t)}{\delta t}.
	\end{equation}

	\section{Experiments}
 	\begin{figure}[t]
		\centering
		\resizebox{0.48\textwidth}{!}
		{\includegraphics{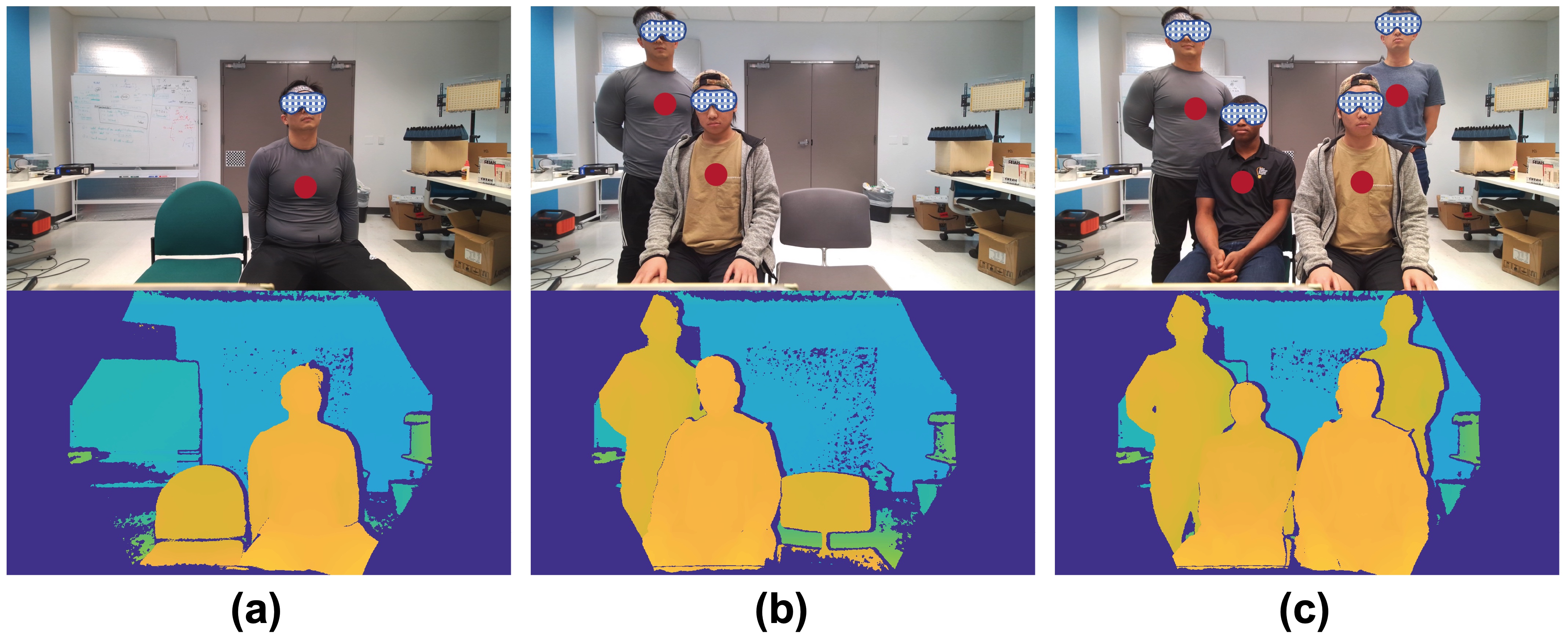}}
		\caption{This figure shows the RGB, torso landmarks, and depth images of the three experiment scenes: (a) single subject; (b) two subjects, seating and standing; (c) a group of four subjects.\label{Fig::Exp::Scenes}}
	\end{figure}

	\begin{figure}[t]
		\centering
		\resizebox{0.48\textwidth}{!}
		{\includegraphics{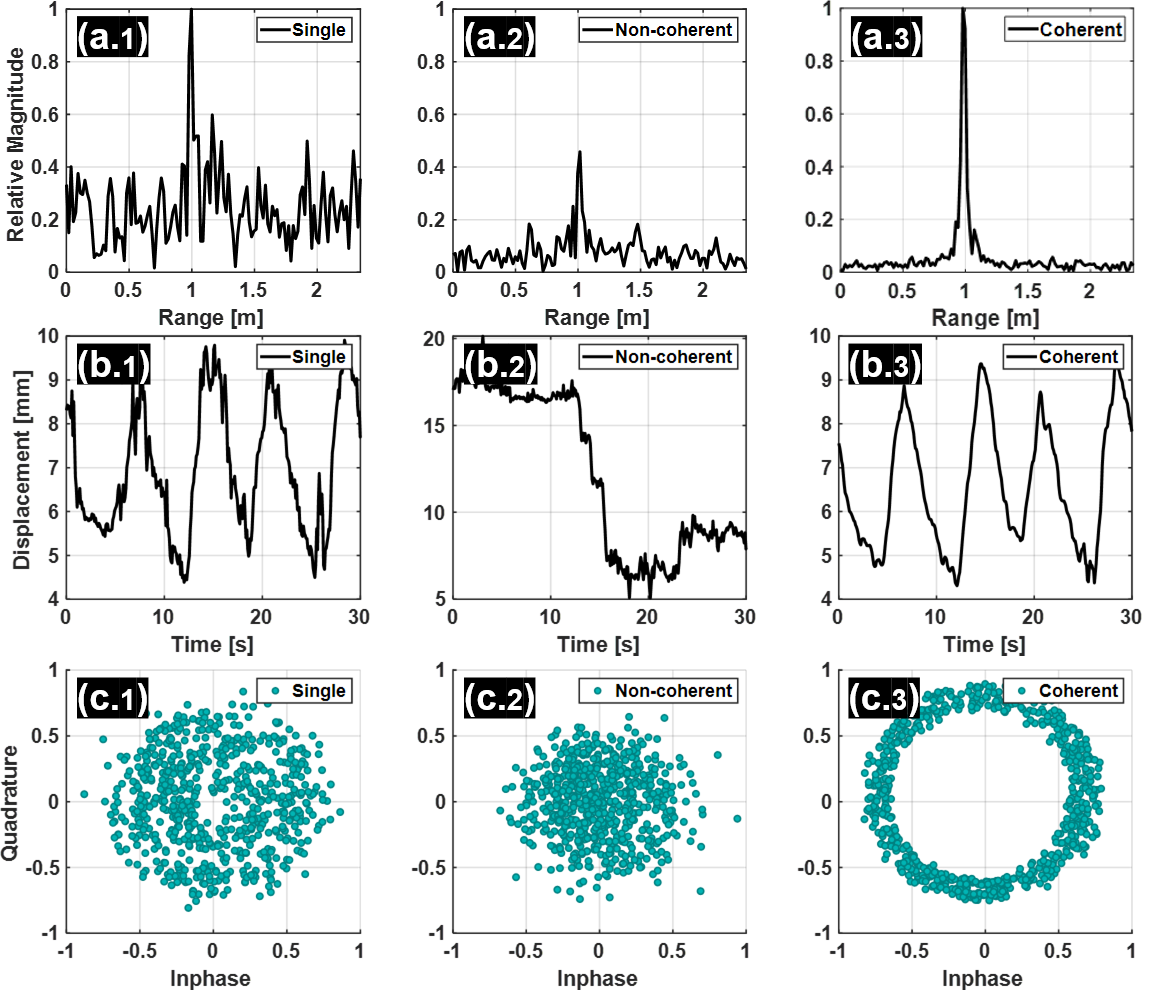}}
		\caption{Signal quality improvement from BF. Top row (a.1) to (a.3) presents range profiles from different processing; middle row (b.1) to (b.3) represents the corresponding respiration patterns from the detected ranges; lower row (c.1) to (c.3) represents the corresponding phase signals in inphase and quadrature plots; from left to right, the columns represent results from single channel, non-coherent combining all channels and coherent combining.}\label{Fig::Exp::Performance}
	\end{figure}
	
	\begin{figure}[t]
		\centering
		\resizebox{0.48\textwidth}{!}
		{\includegraphics{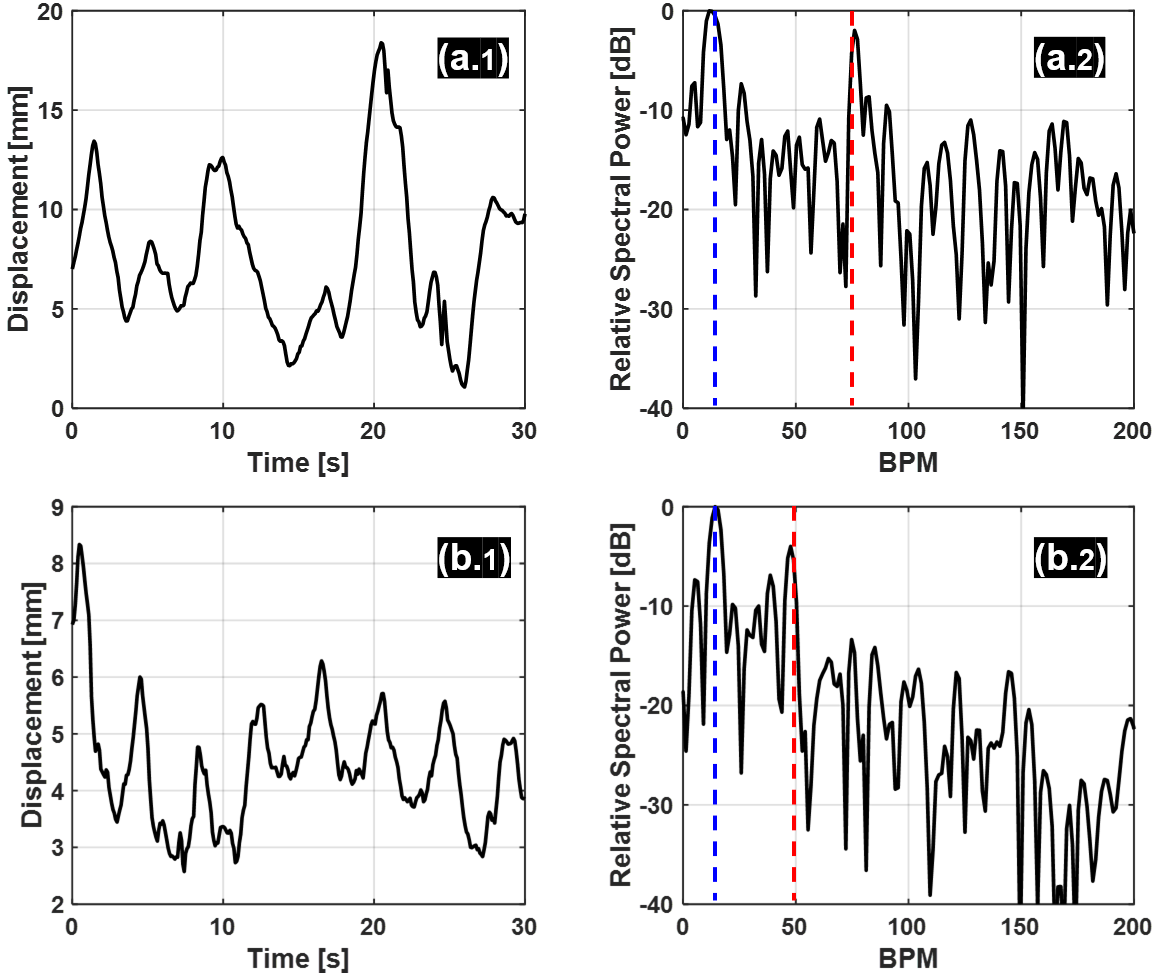}}
		\caption{Measurements from two subjects experiment in Fig. \ref{Fig::Exp::Scenes}(2). (a.1) and (a.2) denote the estimated respiration pattern and the vital signs spectrum from the standing subject while (b.1) and (b.2) from the seating subject. The dashed blue and red lines are the BR and HR derived from the breathing belt and the oximeter.}\label{Fig::Exp2}
	\end{figure}

	\begin{figure}[t]
		\centering
		\resizebox{0.48\textwidth}{!}
		{\includegraphics{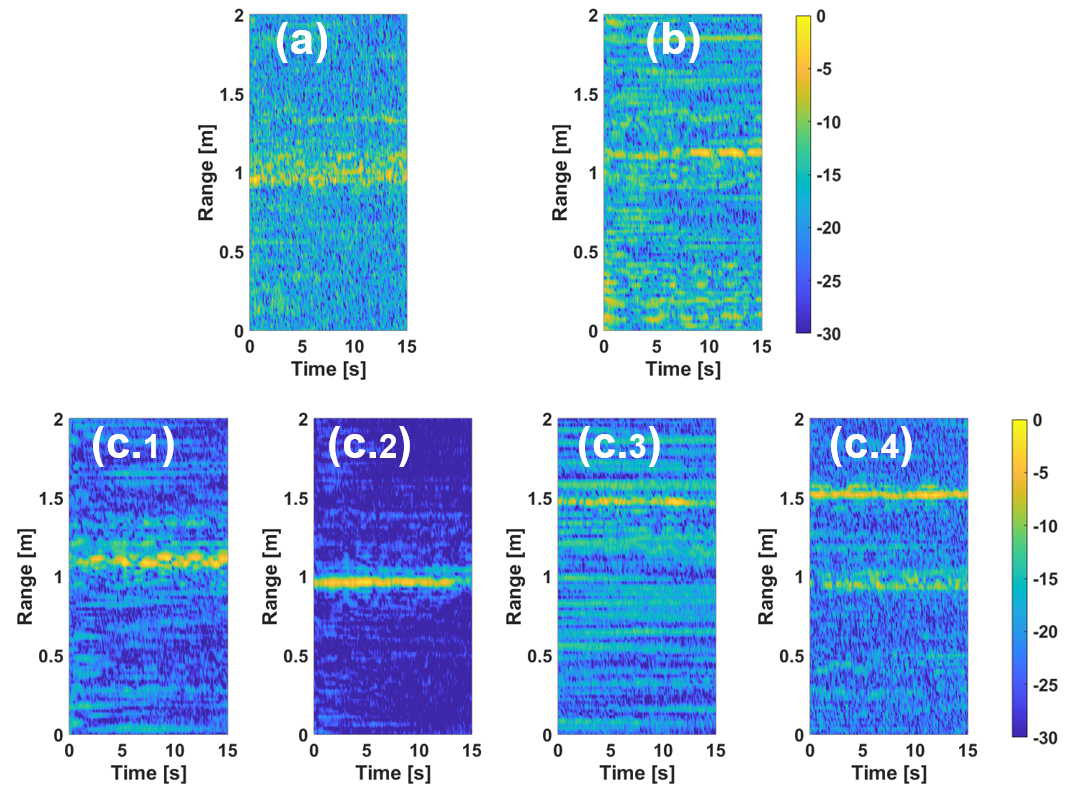}}
		\caption{Range-time heatmap comparison after BF for experiment in Fig. \ref{Fig::Exp::Scenes}(3). (a) single channel; (b) non-coherent combining all channels and (c.1)-(c.4) vision aided BF for separating four subjects. All results are normalized and the color map are represented in dB power.}\label{Fig::Exp3::RT}
	\end{figure}
	
	\begin{figure}[t]
		\centering
		\resizebox{0.48\textwidth}{!}
		{\includegraphics{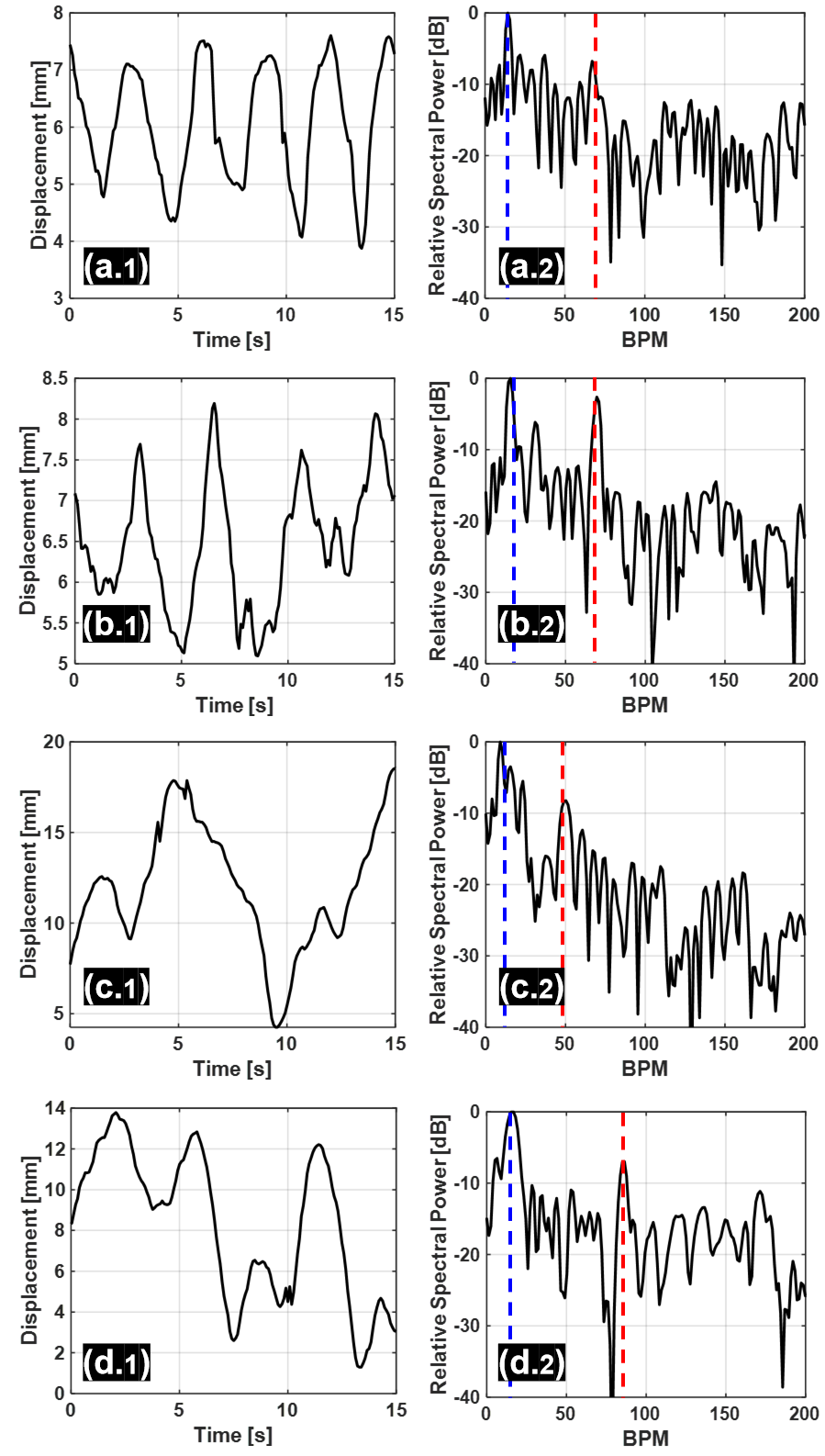}}
		\caption{VSD results for experiment in Fig. \ref{Fig::Exp::Scenes}(3). (a.1) and (a.2) represent the estimated respiration pattern and the vital signs spectrum for the lower-left subject in Fig. \ref{Fig::Exp::Scenes}(3); (b.1) and (b.2) the lower-right subject; (c.1) and (c.2) the top-left subject; (d.1) and (d.2) the top-right subject. The dashed blue and red curves denote the reference BR and HR.}\label{Fig::Exp3::Results}
	\end{figure}

	\subsection{Result Analysis}
	
	DSS is validated in three human tests shown in Fig. \ref{Fig::Exp::Scenes}. Signal quality is inspected at the various stages in the processing chain in the first experiment Fig. \ref{Fig::Exp::Scenes}(1). Two human subjects separated in elevation direction is demonstrated in Fig. \ref{Fig::Exp::Scenes}(2), in which one subject is standing right behind a seating subject. Lastly, VSD from a group of four closely spaced subjects is performed in the last experiment Fig. \ref{Fig::Exp::Scenes}(3).
	
	{\underline{\textit{Signal Enhancement}}}: Constructively combining the spatial channels significantly improves the signal quality in various forms. Top row (a.1) to (a.3) of Fig. \ref{Fig::Exp::Performance} compares the one snapshot of  the range profile from different data processing, single channel only, non-coherent summing all spatial channels and coherent combining. clearly, the range profile from coherent processing gives the best peak detection performance. The naive summation of the spatial channels even degrades the range profile as the subject is at a tilted angle. Similar results are observed in the middle row (b.1) to (b.2). The estimated respiration pattern after BF is less noisy than the single channel result while (b.2) produces an unmeaningful signal pattern. Another way to inspect signal quality is to visualize the inphase and quadrature components. The phase noise performance is dramatically improved as shown in Fig. \ref{Fig::Exp::Performance} shown as a circular arc (c.3) while (c.1) and (c.2) give noise clouds.  
		

	\begin{table*}[h]
		\centering
		\caption{Comparison to State of the Art Methods } \label{TB::Comparison}
		\resizebox{\textwidth}{!}{
			\begin{tabular}{cc|c|c|c|c|c|c|c|c|c|c|}
				\cline{3-12}
				\multicolumn{1}{l}{}              & \multicolumn{1}{l|}{}                                  & \textbf{Domain}                                                       & \textbf{Radar}                                          & \textbf{Arrays}           & \textbf{Camera}              & \textbf{Methodology}                                                & \textbf{No. of Subjects/Posture}                                           & \textbf{Automation} & \textbf{Localization}                                                                                    & \textbf{Vital Signs}                                            & \textbf{Robustness} \\ \hline
				\multicolumn{1}{|c|}{\textbf{1}}  & Rong etc \cite{rong2019smart}                          & Frequency                                                             & \begin{tabular}[c]{@{}c@{}}7.3 GHz \\ UWB\end{tabular}  & SISO                      & /                            & Spectral harmonics                                                  & \begin{tabular}[c]{@{}c@{}}Two, \\ closely spaced, \\ seating \end{tabular}     & No                  & /                                                                                                        & HR                                                              & No                  \\ \hline
				\multicolumn{1}{|c|}{\textbf{2}}  & Rong etc \cite{rong2019smart}                          & Range                                                                 & \begin{tabular}[c]{@{}c@{}}7.3 GHz \\ UWB\end{tabular}  & SISO                      & NA                           & Range resolution                                                    & \begin{tabular}[c]{@{}c@{}}Two, \\ widely separated, \\ seating \end{tabular}   & No                  & NA                                                                                                       & \begin{tabular}[c]{@{}c@{}}RR, \\ HR\end{tabular}               & No                  \\ \hline
				\multicolumn{1}{|c|}{\textbf{3}}  & Xu etc \cite{xu2022simultaneous}                       & \begin{tabular}[c]{@{}c@{}}Azimuth\\ /Range\end{tabular}              & \begin{tabular}[c]{@{}c@{}}77 GHz \\ FMCW\end{tabular}  & 3 x 4                     & NA                           & 1-D BF                                                              & \begin{tabular}[c]{@{}c@{}}Two, \\ widely separated, \\ seating \end{tabular}   & No                  & NA                                                                                                       & \begin{tabular}[c]{@{}c@{}}RR, \\ HR\end{tabular}               & No                  \\ \hline
				\multicolumn{1}{|c|}{\textbf{4}}  & Han etc \cite{han2021detection}                        & \begin{tabular}[c]{@{}c@{}}Azimuth\\ /Range\end{tabular}              & \begin{tabular}[c]{@{}c@{}}77 GHz \\ FMCW\end{tabular}  & 3 x 4                     & NA                           & 1-D BF                                                              & \begin{tabular}[c]{@{}c@{}}Three, \\ widely separated, \\ seating \end{tabular} & No                  & NA                                                                                                       & /                                                               & No                  \\ \hline
				\multicolumn{1}{|c|}{\textbf{5}}  & Mercuri etc \cite{mercuri2022automatic}                & \begin{tabular}[c]{@{}c@{}}Azimuth\\ /Range\end{tabular}              & \begin{tabular}[c]{@{}c@{}}60 GHz \\ FMCW\end{tabular}  & 1 x 4                     & NA                           & \begin{tabular}[c]{@{}c@{}}Eigenvalue \\ decomposition\end{tabular} & \begin{tabular}[c]{@{}c@{}}Two, \\ widely separated, \\ seating \end{tabular}   & Yes                 & \begin{tabular}[c]{@{}c@{}}Z $\sim$ 0.12 m,  \\ Azimuth $\sim$ 3.06\end{tabular}                         & /                                                               & No                  \\ \hline
				\multicolumn{1}{|c|}{\textbf{6}}  & Yan etc \cite{yan2019phase}                            & \begin{tabular}[c]{@{}c@{}}Azimuth\\ /Range\end{tabular}              & \begin{tabular}[c]{@{}c@{}}5.8 GHz \\ FMCW\end{tabular} & SISO                      & NA                           & SAR imaging                                                         & \begin{tabular}[c]{@{}c@{}}Two, \\ widely separated, \\ seating \end{tabular}   & No                  & X/Z $\sim$0.1 m                                                                                          & RR                                                              & No                  \\ \hline
				\multicolumn{1}{|c|}{\textbf{7}}  & Gu etc \cite{gu2013hybrid}                             & NA                                                                    & \begin{tabular}[c]{@{}c@{}}5.8 GHz \\ CW\end{tabular}   & SISO                      & RGB                          & Dual system fusion                                                  & \begin{tabular}[c]{@{}c@{}}Single, \\ seating \end{tabular}                                                            & No                  & NA                                                                                                       & \begin{tabular}[c]{@{}c@{}}RR, \\ HR\end{tabular}               & Limited             \\ \hline
				\multicolumn{1}{|c|}{\textbf{8}}  & Shokouhmand etc \cite{shokouhmand2022camera}           & \begin{tabular}[c]{@{}c@{}}Azimuth\\ /Range\end{tabular}              & \begin{tabular}[c]{@{}c@{}}77 GHz \\ FMCW\end{tabular}  & 3 x 4                     & 3-D depth                    & Vision aided 2-D BF                                                 & \begin{tabular}[c]{@{}c@{}}Two, \\ widely separated, \\ seating \end{tabular}   & Yes                 & NA                                                                                                       & \begin{tabular}[c]{@{}c@{}}RR, \\ HR\end{tabular}               & No                  \\ \hline
				\multicolumn{1}{|l|}{\textbf{9}}  & Chian etc \cite{chian2022vital}                        & NA                                                                    & \begin{tabular}[c]{@{}c@{}}2.4 GHz\\ CW\end{tabular}    & \multicolumn{1}{l|}{SISO} & \multicolumn{1}{l|}{Thermal} & \begin{tabular}[c]{@{}c@{}}Fusion,\\ data mining\end{tabular}       & \begin{tabular}[c]{@{}c@{}}More than two, \\ closely spaced, \\ seating \end{tabular}    & Yes                 & NA                                                                                                       & \begin{tabular}[c]{@{}c@{}}Temperature,\\ RR,\\ HR\end{tabular} & No                  \\ \hline
				\multicolumn{1}{|c|}{\textbf{10}} & This Work                                              & \begin{tabular}[c]{@{}c@{}}Azimuth\\ /Elevation\\ /Range\end{tabular} & \begin{tabular}[c]{@{}c@{}}62 GHz \\ SFCW\end{tabular}  & 20 x 20                   & 3-D depth                    & Vision aided 2-D BF                                                 & \begin{tabular}[c]{@{}c@{}}More than two, \\ closely spaced, \\ seating/standing \end{tabular}    & Yes                 & \begin{tabular}[c]{@{}c@{}}X $\sim$ 0.02 m,  \\ Y $\sim$ 0.04 m, \\ Z $\sim$ 0.005 m \end{tabular}   & \begin{tabular}[c]{@{}c@{}}RR, \\ HR\end{tabular}               & Yes                 \\ \hline
		\end{tabular}}
	\end{table*}

	{\underline{\textit{Elevation Separation}}}: To the best of the authors' knowledge, this is the first study on separating closely spaced subjects in elevation direction using wireless signals and thus recovering vital signs from each individual. The estimated respiration patterns from the two subjects are displayed in the first column of Fig. \ref{Fig::Exp2}. The corresponding spectra are given in the second column. The seating subject produces a slow but irregular respiration pattern (a.1) while the standing subject shows a fast regular respiration pattern (b.1). This observation matches with the RR estimate. Slow breath leads to low RR about 12 BPM while fast breath leads to high RR about 17 BPM. The two subjects also exhibit very different rest HR. They are 78 BPM for the seating subject and 49 BPM for the standing subject. The rest HR estimation is produced from a 15-second data and the estimation is consistent with the HR reference derived from the oximeter. The vast difference in resting HR depends on the subject's physical condition, gender, height, and lifestyle. 
	
	{\underline{\textit{Group of Four Subjects}}}: The last demonstration proves the working of the proposed DSS in crowded space. To visually inspect the isolation performance of a group of subjects, the range-time heatmaps from each individual are displayed in Fig. \ref{Fig::Exp3::RT} (c.1)-(c.4) in comparison with the single channel (a) and non-coherent combining (b). The scene of the experiment is shown in Fig. \ref{Fig::Exp::Scenes}(3). (c.1)-(c.4) are obtained by steering radar BF towards the vision guided directions. The four subjects are located at different radial distances to the system. The strongest color strips in (c.1)-(c.4) indicate the target locations. They occur at approximately 1.1 m in (c.1), 0.98 m in (c.2), 1.47 m in (c.3), and 1.54 m in (c.4). In comparison, it is impossible to infer multiple subjects in (a) and (b) as they are closely spaced and mixed with each other showing only faded strips. Additionally, the dynamic range is much wider in (c.1)-(c.4) in comparison with (a) and (b). Closer subjects enjoy stronger signal return and thus higher dynamic range in (c.2).

	The estimated vital signs from the four subjects are plotted in Fig. \ref{Fig::Exp3::Results}. The first column displays the distinctive respiration patterns and the second column shows the vital signs spectra with the RR and HR highlighted. All the spectral HR signatures are identifiable. The estimated values are within 2 BPM error from the reference values. In particular, the estimated HRs for lower-left subject, lower-right subject, top-left subject, top-right subject are 63 BPM, 78 BPM, 50 BPM and 85 BPM while the reference HRs are 64 BPM, 77 BPM, 48 BPM and 85 BPM, respectively.

	\subsection{Comparison with State-of-the-Art Methods}
	
	A comparison study between our method and other recently developed methods is provided. The main consideration is whether the proposed system can be generalized to practical environments. One typical scene is a group of people in a tight office. This is challenging for the radar itself to automatically identify the number of subjects and also separate vital sign signatures from nearby subjects. For radar-only systems, the vital sign Doppler is often used to achieve human subject identification and localization in the Doppler enhanced range-angle map \cite{han2021detection,mercuri2022automatic,yan2019phase}. Additionally, most radar based methods required knowledge of the number of subjects. A recent study \cite{mercuri2022automatic} exploited eigenvalue decomposition and signal subspace estimation for widely separated human subjects. However, when two subjects are closely spaced, both the steer vectors and the human vital signs are statistically similar and thus the decomposed signals from the radar return are highly correlated. This approach will underestimate the number of subjects. By introducing computer vision, the radar sensing capability can be significantly improved. This is evidenced in \cite{gu2013hybrid} but the dual system was limited to single-subject scenario. The concept of fusion camera motion information and radar signal to cancel random body motion is of profound interest and can be readily incorporated into our system to enable robust sensing of free-living subjects. The most related work on mapping 3-D depth camera and MIMO radar with limited array size \cite{shokouhmand2022camera} was not able to separate multiple closely spaced subjects. The alignment method between camera and radar only considered linear translations and thus cannot be generalized for rotational offsets. 
	
	The overall comparison under various factors are summarized in Table \ref{TB::Comparison}.
	A fair comparison of VSD from these prior works is difficult because VSD performance is highly depends on the Tx power, number of subjects, perspective angle, distance, and processing window length. 
    As a concrete example, the maximum radar Tx power of the developed system is -10 dBm while the systems adopted in the list have much higher Tx power as high as 12 dBm. 
    Another evidence is that the coherent processing window from these works varied from 12 s to 50 s. In that regard, only the types of measurement were listed, such as RR and HR, in our VSD performance comparison.

	\section{Conclusion}
	
	In this paper, we presented a dual-sensor system in which we leveraged computer vision techniques to significantly enhance radar sensing performance in crowded environments. To support the system development, we developed a calibration routine to align the 3-D coordinates from the two imaging modalities. We implemented VSD algorithms, which automatically localized multiple subjects and extracted their vitals. The utilization of two remote sensing technologies, vision plus wireless, opens up numerous potential applications for intelligent healthcare. One possible future application is to investigate the fusion of camera and radar to monitor vital signs of multiple moving subjects.


	\balance
	\bibliographystyle{IEEEtran}

\begin{thebibliography}{10}
	\providecommand{\url}[1]{#1}
	\csname url@samestyle\endcsname
	\providecommand{\newblock}{\relax}
	\providecommand{\bibinfo}[2]{#2}
	\providecommand{\BIBentrySTDinterwordspacing}{\spaceskip=0pt\relax}
	\providecommand{\BIBentryALTinterwordstretchfactor}{4}
	\providecommand{\BIBentryALTinterwordspacing}{\spaceskip=\fontdimen2\font plus
		\BIBentryALTinterwordstretchfactor\fontdimen3\font minus
		\fontdimen4\font\relax}
	\providecommand{\BIBforeignlanguage}[2]{{%
			\expandafter\ifx\csname l@#1\endcsname\relax
			\typeout{** WARNING: IEEEtran.bst: No hyphenation pattern has been}%
			\typeout{** loaded for the language `#1'. Using the pattern for}%
			\typeout{** the default language instead.}%
			\else
			\language=\csname l@#1\endcsname
			\fi
			#2}}
	\providecommand{\BIBdecl}{\relax}
	\BIBdecl
	
	\bibitem{gu2016short}
	C.~Gu, ``Short-range noncontact sensors for healthcare and other emerging
	applications: A review,'' \emph{Sensors}, vol.~16, no.~8, p. 1169, 2016.
	
	\bibitem{mcduff2015survey}
	D.~J. McDuff, J.~R. Estepp, A.~M. Piasecki, and E.~B. Blackford, ``A survey of
	remote optical photoplethysmographic imaging methods,'' in \emph{2015 37th
		annual international conference of the IEEE engineering in medicine and
		biology society (EMBC)}.\hskip 1em plus 0.5em minus 0.4em\relax IEEE, 2015,
	pp. 6398--6404.
	
	\bibitem{magdalena2018sparseppg}
	E.~Magdalena~Nowara, T.~K. Marks, H.~Mansour, and A.~Veeraraghavan,
	``Sparseppg: Towards driver monitoring using camera-based vital signs
	estimation in near-infrared,'' in \emph{Proceedings of the IEEE conference on
		computer vision and pattern recognition workshops}, 2018, pp. 1272--1281.
	
	\bibitem{kempfle2018respiration}
	J.~Kempfle and K.~Van~Laerhoven, ``Respiration rate estimation with depth
	cameras: An evaluation of parameters,'' in \emph{Proceedings of the 5th
		international Workshop on Sensor-based Activity Recognition and Interaction},
	2018, pp. 1--10.
	
	\bibitem{rongt2020respiration}
	Y.~Rong, S.~Srinivas, H.~Chu, H.~Yu, K.~Liu, and D.~W. Bliss, ``Respiration and
	cardiac activity sensing using 3-d cameras,'' in \emph{2020 54th Asilomar
		Conference on Signals, Systems, and Computers}.\hskip 1em plus 0.5em minus
	0.4em\relax IEEE, 2020, pp. 955--959.
	
	\bibitem{chen2006micro}
	V.~C. Chen, F.~Li, S.-S. Ho, and H.~Wechsler, ``Micro-doppler effect in radar:
	phenomenon, model, and simulation study,'' \emph{IEEE Transactions on
		Aerospace and electronic systems}, vol.~42, no.~1, pp. 2--21, 2006.
	
	\bibitem{li2008random}
	C.~Li and J.~Lin, ``Random body movement cancellation in doppler radar vital
	sign detection,'' \emph{IEEE Transactions on Microwave Theory and
		Techniques}, vol.~56, no.~12, pp. 3143--3152, 2008.
	
	\bibitem{rong2018harmonics}
	Y.~Rong and D.~W. Bliss, ``Harmonics-based multiple heartbeat detection at
	equal distance using uwb impulse radar,'' in \emph{2018 IEEE Radar Conference
		(RadarConf18)}.\hskip 1em plus 0.5em minus 0.4em\relax IEEE, 2018, pp.
	1101--1105.
	
	\bibitem{bliss2003multiple}
	D.~Bliss and K.~Forsythe, ``Multiple-input multiple-output (mimo) radar and
	imaging: degrees of freedom and resolution,'' in \emph{The Thrity-Seventh
		Asilomar Conference on Signals, Systems \& Computers, 2003}, vol.~1.\hskip
	1em plus 0.5em minus 0.4em\relax IEEE, 2003, pp. 54--59.
	
	\bibitem{bliss2006mimo}
	------, ``Mimo radar medical imaging: self-interference mitigation for breast
	tumor detection,'' in \emph{2006 Fortieth Asilomar Conference on Signals,
		Systems and Computers}.\hskip 1em plus 0.5em minus 0.4em\relax IEEE, 2006,
	pp. 1558--1562.
	
	\bibitem{rong2019radar}
	Y.~Rong and D.~W. Bliss, ``Is radar cardiography (rcg) possible?'' in
	\emph{2019 IEEE Radar Conference (RadarConf)}.\hskip 1em plus 0.5em minus
	0.4em\relax IEEE, 2019, pp. 1--6.
	
	\bibitem{nosrati2019concurrent}
	M.~Nosrati, S.~Shahsavari, S.~Lee, H.~Wang, and N.~Tavassolian, ``A concurrent
	dual-beam phased-array doppler radar using mimo beamforming techniques for
	short-range vital-signs monitoring,'' \emph{IEEE Transactions on Antennas and
		Propagation}, vol.~67, no.~4, pp. 2390--2404, 2019.
	
	\bibitem{feng2021multitarget}
	C.~Feng, X.~Jiang, M.-G. Jeong, H.~Hong, C.-H. Fu, X.~Yang, E.~Wang, X.~Zhu,
	and X.~Liu, ``Multitarget vital signs measurement with chest motion imaging
	based on mimo radar,'' \emph{IEEE Transactions on Microwave Theory and
		Techniques}, vol.~69, no.~11, pp. 4735--4747, 2021.
	
	\bibitem{ahmad2018vital}
	A.~Ahmad, J.~C. Roh, D.~Wang, and A.~Dubey, ``Vital signs monitoring of
	multiple people using a fmcw millimeter-wave sensor,'' in \emph{2018 IEEE
		Radar Conference (RadarConf18)}.\hskip 1em plus 0.5em minus 0.4em\relax IEEE,
	2018, pp. 1450--1455.
	
	\bibitem{rong2021radar}
	Y.~Rong, K.~V. Mishra, and D.~W. Bliss, ``Radar-based radial arterial pulse
	rate and pulse pressure analysis,'' in \emph{2021 29th European Signal
		Processing Conference (EUSIPCO)}.\hskip 1em plus 0.5em minus 0.4em\relax
	IEEE, 2021, pp. 1870--1874.
	
	\bibitem{rong2020cardiac}
	Y.~Rong, P.~C. Theofanopoulos, G.~C. Trichopoulos, and D.~W. Bliss, ``Cardiac
	sensing exploiting an ultra-wideband terahertz sensing system,'' in
	\emph{2020 IEEE International Radar Conference (RADAR)}.\hskip 1em plus 0.5em
	minus 0.4em\relax IEEE, 2020, pp. 1002--1006.
	
	\bibitem{gu2013hybrid}
	C.~Gu, G.~Wang, Y.~Li, T.~Inoue, and C.~Li, ``A hybrid radar-camera sensing
	system with phase compensation for random body movement cancellation in
	doppler vital sign detection,'' \emph{IEEE transactions on microwave theory
		and techniques}, vol.~61, no.~12, pp. 4678--4688, 2013.
	
	\bibitem{chian2022vital}
	D.-M. Chian, C.-K. Wen, C.-J. Wang, M.-H. Hsu, and F.-K. Wang, ``Vital signs
	identification system with doppler radars and thermal camera,'' \emph{IEEE
		Transactions on Biomedical Circuits and Systems}, vol.~16, no.~1, pp.
	153--167, 2022.
	
	\bibitem{shokouhmand2022camera}
	A.~Shokouhmand, S.~Eckstrom, B.~Gholami, and N.~Tavassolian, ``Camera-augmented
	non-contact vital sign monitoring in real time,'' \emph{IEEE Sensors
		Journal}, 2022.
	
	\bibitem{xu2022simultaneous}
	Z.~Xu, C.~Shi, T.~Zhang, S.~Li, Y.~Yuan, C.-T.~M. Wu, Y.~Chen, and
	A.~Petropulu, ``Simultaneous monitoring of multiple people’s vital sign
	leveraging a single phased-mimo radar,'' \emph{IEEE Journal of
		Electromagnetics, RF and Microwaves in Medicine and Biology}, 2022.
	
	\bibitem{xiong2021millimeter}
	Y.~Xiong, S.~Li, C.~Gu, G.~Meng, and Z.~Peng, ``Millimeter-wave bat for mapping
	and quantifying micromotions in full field of view,'' \emph{Research}, vol.
	2021, 2021.
	
	\bibitem{turaga2008machine}
	P.~Turaga, R.~Chellappa, V.~S. Subrahmanian, and O.~Udrea, ``Machine
	recognition of human activities: A survey,'' \emph{IEEE Transactions on
		Circuits and Systems for Video technology}, vol.~18, no.~11, pp. 1473--1488,
	2008.
	
	\bibitem{sarafianos20163d}
	N.~Sarafianos, B.~Boteanu, B.~Ionescu, and I.~A. Kakadiaris, ``3d human pose
	estimation: A review of the literature and analysis of covariates,''
	\emph{Computer Vision and Image Understanding}, vol. 152, pp. 1--20, 2016.
	
	\bibitem{noon1996stepped}
	D.~A. Noon, ``Stepped-frequency radar design and signal processing enhances
	ground penetrating radar performance,'' 1996.
	
	\bibitem{rong2022vital}
	Y.~Rong, I.~Lenz, and D.~W. Bliss, ``Vital signs detection based on
	high-resolution 3-d mmwave radar imaging,'' in \emph{2022 IEEE International
		Symposium on Phased Array Systems \& Technology (PAST)}.\hskip 1em plus 0.5em
	minus 0.4em\relax IEEE, 2022, pp. 1--6.
	
	\bibitem{nikolova2017introduction}
	N.~K. Nikolova, \emph{Introduction to microwave imaging}.\hskip 1em plus 0.5em
	minus 0.4em\relax Cambridge University Press, 2017.
	
	\bibitem{bamji2018impixel}
	C.~S. Bamji, S.~Mehta, B.~Thompson, T.~Elkhatib, S.~Wurster, O.~Akkaya,
	A.~Payne, J.~Godbaz, M.~Fenton, V.~Rajasekaran \emph{et~al.}, ``Impixel 65nm
	bsi 320mhz demodulated tof image sensor with 3$\mu$m global shutter pixels
	and analog binning,'' in \emph{2018 IEEE International Solid-State Circuits
		Conference-(ISSCC)}.\hskip 1em plus 0.5em minus 0.4em\relax IEEE, 2018, pp.
	94--96.
	
	\bibitem{azureDK_spec}
	\BIBentryALTinterwordspacing
	{Microsoft, Inc}. (2020, Feb) {Azure Kinect DK hardware specifications}.
	[Online]. Available:
	\url{https://docs.microsoft.com/en-us/azure/kinect-dk/hardware-specification}
	\BIBentrySTDinterwordspacing
	
	\bibitem{Pose}
	\BIBentryALTinterwordspacing
	{Google, Inc}, ``{MediaPipe Pose},'' Nov. 2022. [Online]. Available:
	\url{https://google.github.io/mediapipe/solutions/pose.html}
	\BIBentrySTDinterwordspacing
	
	\bibitem{rong2019smart}
	Y.~Rong and D.~W. Bliss, ``Smart homes: See multiple heartbeats through wall
	using wireless signals,'' in \emph{2019 IEEE Radar Conference
		(RadarConf)}.\hskip 1em plus 0.5em minus 0.4em\relax IEEE, 2019, pp. 1--6.
	
	\bibitem{han2021detection}
	K.~Han and S.~Hong, ``Detection and localization of multiple humans based on
	curve length of i/q signal trajectory using mimo fmcw radar,'' \emph{IEEE
		Microwave and Wireless Components Letters}, vol.~31, no.~4, pp. 413--416,
	2021.
	
	\bibitem{mercuri2022automatic}
	M.~Mercuri, P.~Russo, M.~Glassee, I.~D. Castro, E.~De~Greef, M.~Rykunov,
	M.~Bauduin, A.~Bourdoux, I.~Ocket, F.~Crupi \emph{et~al.}, ``Automatic
	radar-based 2-d localization exploiting vital signs signatures,''
	\emph{Scientific Reports}, vol.~12, no.~1, pp. 1--11, 2022.
	
	\bibitem{yan2019phase}
	J.~Yan, G.~Zhang, H.~Hong, H.~Chu, C.~Li, and X.~Zhu, ``Phase-based human
	target 2-d identification with a mobile fmcw radar platform,'' \emph{IEEE
		Transactions on Microwave Theory and Techniques}, vol.~67, no.~12, pp.
	5348--5359, 2019.
	
\end{thebibliography}

	\vfill
	
\end{document}